\documentclass[preprint,12pt]{emulateapj}
\usepackage{color}
\usepackage{natbib}
\usepackage{amssymb,amsmath}
\usepackage{graphicx}
\usepackage{multirow}
\usepackage[pdftex,                
bookmarks=true,                    
bookmarksnumbered=true,            
colorlinks=true,            
citecolor=blue,         
linkcolor=blue,     
menucolor=blue,         
urlcolor=cyan,       
linkbordercolor={0 0 1},           
pdfborder= {0 0 1},
frenchlinks=True]{hyperref}

\providecommand{\eprint}[1]{\href{http://arxiv.org/abs/#1}{#1}}
\providecommand{\adsurl}[1]{\href{#1}{ADS}}

\newcommand{\hdoneb}{HD~189733b}

\def\aap{{A\&A}}		
\def\apj{{ApJ}}			
\def\apjl{{ApJ}}		
\def\pasp{{PASP}}		

\def\mnras{{MNRAS}}
\def\nat{{Nature}}

\def\methane{\ensuremath{\textrm{CH}_4}}
\def\coo{\ensuremath{\textrm{CO}_2}}

\def\deg{\ensuremath{^\circ}}

\def\hkcolorfirst{\ensuremath{0.31\,\% \pm 0.69\,\%}}
\def\hkcolor{\ensuremath{0.137\,\% \pm 0.054\,\%}}
\def\jkcolor{\ensuremath{0.36\,\% \pm 0.14\,\%}}
\def\hktemp{\ensuremath{2400^{+1500}_{-500}~\textrm{K}}}
\def\methanedepth{\ensuremath{27~\textrm{nm}}}
\def\emeandif{\ensuremath{0.029\,\%}}
\def\meandif{\ensuremath{0.236\,\% \pm \emeandif}}

\newcommand{\wt}{\mbox{WASP-12}}
\newcommand{\wtb}{\mbox{WASP-12b}}

\long\def\symbolfootnote[#1]#2{\begingroup%
  \def\thefootnote{\fnsymbol{footnote}}\footnote[#1]{#2}\endgroup} 

\newcommand{\fig}[5]{
        \begin{figure}[!bt]
        \begin{center}
        \includegraphics[#3]{#2}
      \end{center}
      
        \renewcommand{\baselinestretch}{1}
        \vspace*{-.3in}
        \caption[#4]{#5}
        \label{fig:#1}
        \end{figure}}

\newcommand{\figtwocol}[5]{
        \begin{figure*}[!bt]
        \begin{center}
        \includegraphics[#3]{#2}
      \end{center}
      \renewcommand{\baselinestretch}{1}
        \vspace*{-.3in}
        \caption[#4]{\footnotesize #5}
        \label{fig:#1}
        \end{figure*}}

\shortauthors{Crossfield, Hansen, and Barman}
\shorttitle{Ground-based NIR Spectroscopy of \wtb}
\begin{document}


\title{Ground-based, Near-infrared Exospectroscopy. II. Tentative
  Detection of Emission From the Extremely Hot Jupiter \wtb }

\slugcomment{Accepted to ApJ: 2012 Jan 04}

\author{
Ian J. M. Crossfield\altaffilmark{1},
Brad M. S. Hansen\altaffilmark{1},
Travis Barman\altaffilmark{2}}

\altaffiltext{1}{Department of Physics \& Astronomy, University of California Los Angeles, Los Angeles, CA 90095, ianc@astro.ucla.edu}
\altaffiltext{2}{Lowell Observatory, 1400 West Mars Hill Road, Flagstaff, AZ 86001, USA}

\begin{abstract}
  We report the tentative detection of the near-infrared emission of
  the Hot Jupiter WASP-12b with the low-resolution prism on IRTF/SpeX.
  We find a $K-H$ contrast color of \hkcolor, corresponding to a
  blackbody of temperature \hktemp\ and consistent with previous,
  photometric observations.  We also revisit \wtb 's energy budget on
  the basis of secondary eclipse observations: the dayside luminosity
  is a relatively poorly constrained $(2.0-4.3) \times
  10^{30}$~erg~s$^{-1}$, but this still allows us to predict a
  day/night effective temperature contrast of $200-1$,000~K (assuming
  $A_B=0$).  Thus we conclude that \wtb\ probably does not have both a
  low albedo and low recirculation efficiency.  Our results show the
  promise and pitfalls of using single-slit spectrographs for
  characterization of extrasolar planet atmospheres, and we suggest
  future observing techniques and instruments which could lead to
  further progress.  Limiting systematic effects include the use of a
  too-narrow slit on one night -- which observers could avoid in the
  future -- and chromatic slit losses (resulting from the variable
  size of the seeing disk) and variations in telluric transparency --
  which observers cannot control. Single-slit observations of the type
  we present remain the best option for obtaining $\lambda >
  1.7\,\micron$ spectra of transiting exoplanets in the brightest
  systems.  Further and more precise spectroscopy is needed to better
  understand the atmospheric chemistry, structure, and energetics of
  this, and other, intensely irradiated planet.
\end{abstract}

\keywords{infrared: stars --- planetary systems --- stars:~individual
  (\wt) --- techniques: spectroscopic}

\section{Introduction}
\subsection{Ground-based Characterization of Exoplanet Atmospheres}
Transiting extrasolar planets allow the exciting possibility of
studying the intrinsic physical properties of these planets. The last
several years have seen rapid strides in this direction, with
measurements of precise masses and radii, detection of numerous
secondary eclipses and phase curves, and the start of ground-based
optical spectroscopy \citep{redfield:2008, snellen:2008,bean:2010}.

Though ground-based, near-infrared (NIR) photometry of exoplanets is
becoming commonplace, until recently there were no successful
detections via ground-based NIR spectroscopy
\citep{brown:2002,richardson:2003a,deming:2005,barnes:2007a,knutson:2007a}. Several
groups have employed high-resolution spectrographs with some form of
template cross-correlation
\citep{deming:2005,snellen:2010,crossfield:2011} with varying degrees
of success. Though cross-correlation provides a method to test for the
detection of a particular model, it has the significant drawback that
it does not provide a model-independent measurement.  Furthermore,
such observations require high-resolution cryogenic spectrographs on
large-aperture telescopes.

The only published, model-independent, ground-based, NIR spectrum of
an exoplanetary atmosphere \citep{swain:2010} was obtained with a
different approach: medium-resolution spectroscopy of HD~189733b with
the 3~m NASA Infrared Telescope Facility (IRTF) covering the K and L
bands.  However, these results are in dispute: the K band matches
HST/NICMOS observations which have in turn been called into question
\citep[see][]{swain:2008,sing:2009haze,gibson:2011,deroo:2010}, while
the L band exhibits an extremely high flux peak attributed variously
to non-LTE \methane\ emission \citep{swain:2010} and to contamination
by telluric water vapor \citep{mandell:2011}.  In contrast, the
tentative spectroscopic detection of \wtb\ we present in this paper
reproduces previous, high S/N ground-based photometry
\citep{croll:2011a} and we demonstrate that our final result is not
likely to be corrupted by telluric variations outside of well-defined
spectral regions.

\subsection{The  \wt\ System}
The transiting Hot Jupiter \wtb\ has an orbital period of 1.1~days
around its 6300~K host star, and the planet's mass and radius give it
a bulk density only 25\% of Jupiter \citep{hebb:2009,chan:2011}.  The
planet is one of the largest known and is significantly overinflated
compared to standard interior models \citep{fortney:2007}.  The planet
is significantly distorted and may be undergoing Roche lobe overflow
\citep{li:2010}, but tidal effects are not expected to be a
significant energy source. Though the initial report suggested \wtb\
had a nonzero eccentricity, subsequent orbital characterization via
timing of secondary eclipses \citep{campo:2011} and further radial
velocity measurements \citep{husnoo:2011} suggest an eccentricity
consistent with zero.

\wtb\ is intensely irradiated by its host star, making the planet one
of the hottest known and giving it a favorable ($\gtrsim 10^{-3}$) NIR
planet/star flux contrast ratio; it has quickly become one of the
best-studied exoplanets.  The planet's large size, low density, and
high temperature have motivated an ensemble of optical,
\citep{lopez-morales:2010}, NIR \citep{croll:2011a}, and mid-infrared
\citep{campo:2011} eclipse photometry which suggests this planet has
an unusual carbon to oxygen (C/O) ratio greater than one
\citep{madhusudhan:2011}.

However, a wide range of fiducial atmospheric models fit \wtb 's
photometric emission spectrum equally well despite differing
significantly in atmospheric abundances and in their
temperature-pressure profiles \citep{madhusudhan:2011}.  Many hot
Jupiters appear to have high-altitude temperature inversions
\citep{knutson:2010, madhusudhan:2010}, but even WASP-12b's precise,
well-sampled photometric spectrum does not constrain the presence or
absence of such an inversion. Thus significant degeneracies remain;
this is a common state of affairs in the field at present even for
such relatively well-characterized systems\
\citep{madhusudhan:2010}. This is because (a) broadband photometry
averages over features caused by separate opacity sources and (b)
atmospheric models have many more free parameters than there are
observational constraints.  Spectroscopy, properly calibrated, can
break some of these degeneracies, test the interpretation of
photometric observations at higher resolution, and ultimately has the
potential to more precisely refine estimates of atmospheric
abundances, constrain planetary temperature structures, and provide
deeper insight into high-temperature exoplanetary atmospheres.

\subsection{Outline}
This paper presents our observations and analysis of two eclipses of
\wtb\ in an attempt to detect and characterize the planet's NIR
emission spectrum.  This is part of our ongoing effort to develop the
methods necessary for robust, repeatable ground-based exoplanet
spectroscopy, and we use many of the same techniques introduced in our
first paper \citep[][hereafter Paper I]{crossfield:2011}.

We describe our spectroscopic observations and initial data reduction
in Sec.~\ref{sec:obs}.  The data exhibit substantial correlated
variability, and we describe our measurements of various instrumental
variations in Sec.~\ref{sec:systematics}.  We fit a simple model that
includes astrophysical, instrumental, and telluric effects to the data
in Sec.~\ref{sec:detection}. Chromatic slit losses (resulting from
wavelength-dependent atmospheric dispersion and seeing) and telluric
transmittance and radiance effects can confound ground-based NIR
observations, so in Sec.~\ref{sec:sys} we investigate these systematic
error sources in detail. We present our main result -- a tentative
detection of \wtb's emission -- in Sec.~\ref{sec:results} and compare
it to previous observations.  Finally, we discuss the implications of
our work for future ground-based, NIR spectroscopy in
Sec.~\ref{sec:future} and conclude in Sec.~\ref{sec:conclusion}.

\section{Observations and Initial Reduction}
\label{sec:obs}

\subsection{Summary of Observations}
We observed the \wt\ system with the SpeX near-infrared spectrograph
\citep{rayner:2003}, mounted at the IRTF Cassegrain focus.  Our
observations during eclipses on 28 and 30~December 2009 (UT) comprise
a total of 10.2~hours on target and 8.3~hours of integration time. We
list the details of our observations and our instrumental setup in
Table~\ref{tab:observations}.  One of our eclipses overlaps one of
those observed by \cite{croll:2011a} with broadband photometry from
the Canada-France-Hawaii Telescope (CFHT), also on Mauna Kea, on UT
27-29~December 2009.  Our first night, 28~Dec, is the same night as
their H~band observation.

\begin{deluxetable}{l c c }
  \tablecolumns{3} \tablecaption{\label{tab:observations} Observations
  } \startdata
  UT date                   & 2009 Dec 28  & 2010 Dec 30 \\
  Instrument Rotator Angle  & 225\deg & 225\deg  \\
  Slit Position Angle & 90\deg & 90\deg \\
  Slit                     & 1.6'' x 15'' & 3.0'' x 15''  \\
  Grating                   & LowRes15 & LowRes15 \\
  Guiding filter            & J   & K  \\
  OS filter                 & open   & open  \\
  Dichroic                  & open   & open \\
  Integration Time (sec)   & 15 & 20 \\
  Non-destructive reads & 4 & 4 \\
  Co-adds        & 2 & 2 \\
  Exposures      & 502 & 356\tablenotemark{a} \\
  Airmass range            & 1.01 - 1.91 & 1.01 - 2.70\tablenotemark{a} \\
  Wavelength coverage ($\mu$m)   &   $<1$ - 2.5\,\micron & $<1$ - 2.5\,\micron  \\
  \wtb\ phase coverage           & 0.42 - 0.61 & 0.41 - 0.61 \\
\enddata
\tablenotetext{a}{We limit the 30~Dec observations to airmass less
  than 2.326, which reduces the number of usable frames from 379 to 356.}
\end{deluxetable}    

On both nights we observed the \wt\ system continuously for as long as
conditions permitted using SpeX's low-resolution prism mode, which
gives uninterrupted wavelength coverage from $<1-2.5$\,\micron.  We
chose prism mode because it offers roughly twice the throughput of to
SpeX's echelle modes \citep[][their Fig.~7]{rayner:2003}, though it
has a necessarily reduced capability to spectrally resolve, separate,
and mitigate telluric features. We nodded the telescope along the slit
to remove the sky background; as we discuss below, this induced
substantial flux variations in our spectrophotometry at shorter
wavelengths and we urge future exoplanet observers to eschew nodding
at these wavelengths (the exception to this rule would be for
instruments that suffer from time-varying scattered light, such as
SpeX's short-wavelength cross-dispersed mode). We deactivated the
instrument's field rotator to minimize instrumental flexure, but this
meant the slit did not track the parallactic angle and atmospheric
dispersion (coupled with variable seeing and telescope guiding errors)
causes large-scale, time-dependent, chromatic gradients throughout the
night.  As we describe in Sec.~\ref{sec:atmodisp}
and~\ref{sec:allsys}, this effect is reduced (but not eliminated) by
using a wider slit, and we strongly advise that future observations
covering a large wavelength range (a) use as large a slit as possible
and (b) keep the slit aligned to the parallactic angle.

On our first night, 28~Dec, we observed with the 1.6'' slit to
strike a balance between sky background and frame-to-frame variations
in the amount of light entering the slit.  After this run an initial
analysis suggested we could further decrease spectrophotometric
variability without incurring significant penalties from sky
background, and so we used the 3.0'' slit on the second night
(30~Dec).

\wt\ is sufficiently bright (K=10.2) that we were able to guide on the
faint ghost reflected from the transmissive, CaF slit mask into the
NIR slit-viewing guide camera.  Guiding kept the K band relatively
stationary but because SpeX covers such a wide wavelength range the
spectra suffer from differential atmospheric refraction; this results
in substantially larger motions over the course of the night at
shorter wavelengths.  We did not record guide camera frames, but we
recommend that future observers save all such data to track guiding
errors, measure the morphology of the two-dimensional point spread
function, and measure the amount of light falling outside the slit.
Typical frames had maximum count rates of
$\lesssim$2,000~ADU~pix$^{-1}$~coadd$^{-1}$, safely within the
$1024^2$ Aladdin~3 InSb detector's linear response range.

\subsection{Initial Data Reduction}
\label{sec:reduction} We reduce the raw echelleograms using the
SpeXTool reduction package \citep{cushing:2004}, supplemented by our
own set of Python analysis tools.  SpeXTool dark-subtracts,
flat-fields, and corrects the recorded data for detector
nonlinearities, and we find it to be an altogether excellent reduction
package that future instrument teams would do well to emulate.  We
used SpeXTool in optimal ``A$-$B'' point source extraction mode with
extraction and aperture radii of 2.5'', inner and outer background
aperture radii of 2.8'' and 3.5'', respectively, and a linear
polynomial to fit and remove the residual background in each column.

The extracted spectra have H and K band fluxes of $\sim$5500 and
$\sim$2000~e$^-$~pix$^{-1}$~s$^{-1}$, respectively.  After removing
observations rendered unusable for telescope or instrumental reasons
(e.g., loss of guiding or server crashes), we are left with 502 and
356 usable frames from our two nights.  The extracted spectra are
shown in Fig.~\ref{fig:rawdata} and substantial variations are
apparent; we discuss these in Sec.~\ref{sec:systematics}.

\figtwocol{rawdata}{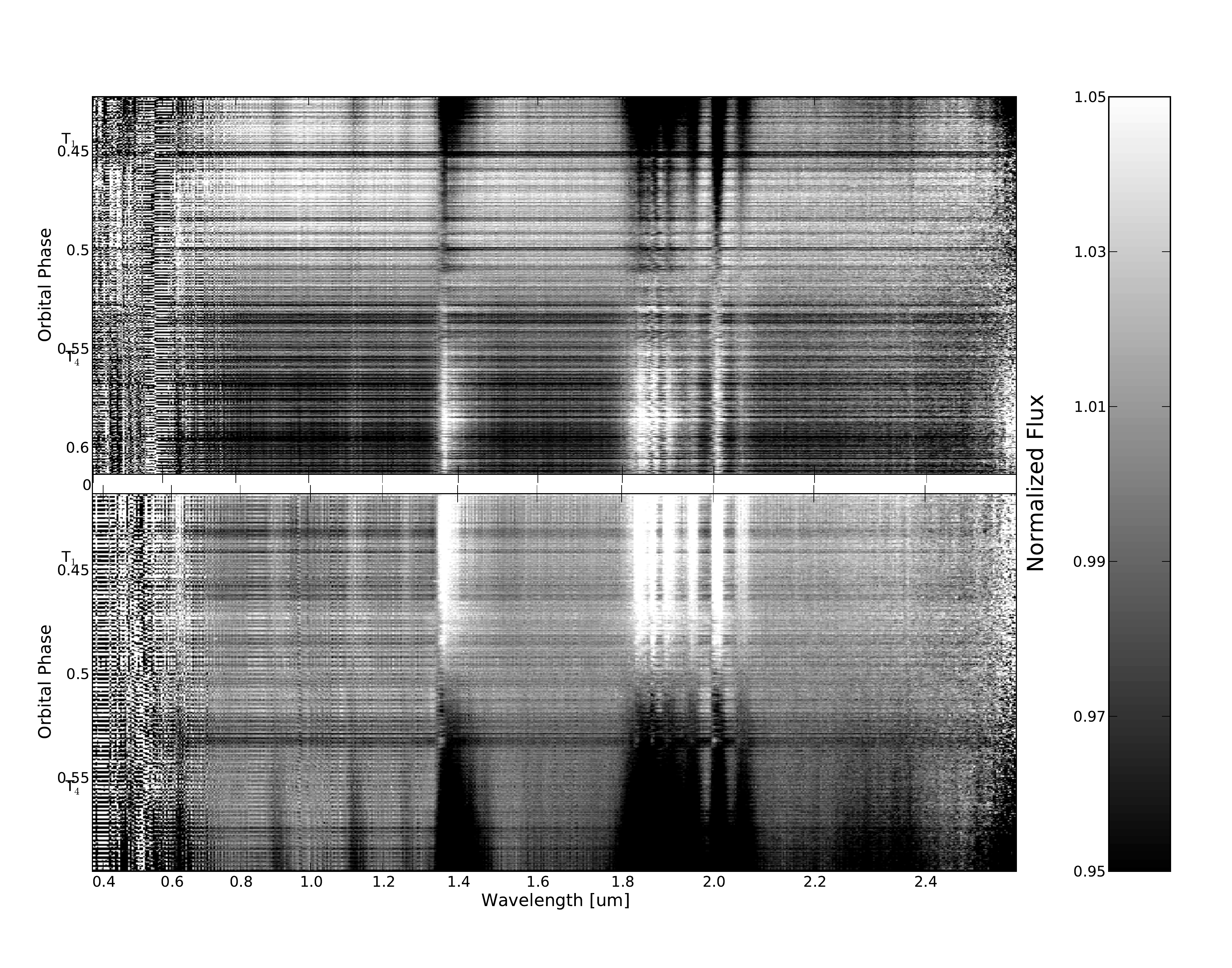}{width=7in}{}{Spectrophotometric data for
  the nights of 28~Dec~2009 ({\em top}) and 30~Dec~2009 ({\em
    bottom}); each column has been normalized by its median value. The
  variations (due to a combination of airmass effects and instrumental
  slit loss) are largely common mode; variations are less on 30~Dec,
  probably because of the wider slit used then.  The first ($T_1$)
  and fourth ($T_4$) points of contact of the eclipse are indicated,
  as calculated from the ephemeris of \cite{hebb:2009}.}

SpeX typically uses a set of arc lamps for wavelength calibration, but
SpeXTool fails to process arcs taken with the 3'' slit in prism mode.
Instead, we calculate wavelength solutions by matching observed
telluric absorption features with an empirical high-resolution
telluric absorption spectrum \citep{hinkle:2003} convolved to the
approximate spectral resolution of our observations.  We estimate a
precision of 1.7~nm for the individual line positions and use this
uncertainty to calculate the $\chi^2$ and Bayesian Information
Criterion\footnote{Bayesian Information Criterion (BIC) = $\chi^2 + k
  \ln N$, where $k$ is the number of free parameters and $N$ the
  number of data points.}  (BIC) for fits using successively higher
degrees of polynomials: for both nights a fourth-order polynomial
gives the lowest BIC, indicating this to be the preferred model.  The
RMS of the residuals to these fits are 1.3 and 1.6~nm for 28~Dec and
30~Dec, respectively, while maximum residuals for each night are
3.1~nm (at 2.35\,\micron) and 2.9~nm (at 1.3, 1.85, and
2.32\,\micron), respectively.


Our wavelength solutions for 28~and 30~Dec are respectively
\begin{eqnarray}
\lambda_{28}(p)/\micron & = & 6.77626638 \times 10^{-12} p^4 - 9.82847002 \times 10^{-9} p^3 + \nonumber \\
& & 2.52383166 \times 10^{-6} p^2 + 4.26945216 \times 10^{-3} p + \nonumber \\
& & 0.414989079 \nonumber
\end{eqnarray}
and 
\begin{eqnarray}
\lambda_{30}(p)/\micron & = & 4.85652782 \times 10^{-12} p^4 - 6.49601098 \times 10^{-9} p^3 + \nonumber \\
& & 4.97530220 \times 10^{-7} p^2 + 4.78270230 \times 10^{-3} p + \nonumber \\
& & 0.371045967 \nonumber
\end{eqnarray}
where $p$ is the pixel number, an integer from 0 to 563, inclusive.
We apply these wavelength solutions to all our spectra after shifting
them to a common reference frame using the shift-and-fit technique
described by \cite{deming:2005} and implemented in Paper~I.

\section{Characterization of Systematic Effects}
\label{sec:systematics}
\subsection{Instrumental Sources}
The initially extracted spectra shown in Fig.~\ref{fig:rawdata}
exhibit temporal variations due to a combination of telluric,
instrumental, and astrophysical sources, with the last of these the
weakest of the three effects.  We wish to quantify and remove the
instrumental and telluric effects to the extent that we can
convincingly detect any astrophysical signature -- i.e., a secondary
eclipse.  The strongest variations in Fig.~\ref{fig:rawdata} are
largely common-mode (i.e., they appear in all wavelength channels) and are due
to variations in light coupled into the spectrograph due to changes in
seeing, pointing, and/or telluric transparency. Longer-term telluric
variations are distinguishable by the manner in which they increase in
severity in regions of known telluric absorption.

We approximate the amount of light coupled into the spectrograph slit
by measuring the flux in regions clear of strong telluric absorption,
as determined using our high-resolution telluric absorption spectrum
\citep{hinkle:2003} convolved to our approximate resolution. The flux
in these channels should only depend on the frame-to-frame changes in
starlight entering the spectrograph slit, which in turn depends on the
(temperature- and pressure-dependent) atmospheric dispersion, the
(wavelength-dependent) size and shape of the instrumental response,
telescope guiding errors, and achromatic changes in telluric
transparency.  In the interests of simplicity we initially treat this
as a wavelength-independent quantity; we return to address the
validity and limitations of this assumption in
Sec.~\ref{sec:atmodisp}.

At each time step we sum the flux in these telluric-free parts of each
spectrum, creating a time series representative of the achromatic slit
losses suffered by the instrument.  Although we refer to this quantity
as the slit loss, it is actually a combination of instrumental slit
losses (spillover) and changing atmospheric transmission.  The
achromatic slit loss time series is plotted for each night in
Figs.~\ref{fig:dec28_state_vectors} and~\ref{fig:dec30_state_vectors},
along with other candidate systematic sources described below.  We
ultimately compute this quantity by summing the flux between
$1.63-1.73\,\micron$ and $2.10-2.21\,\micron$, spectral regions we
show in Sec.~\ref{sec:telluric} to be mostly free of telluric
contamination.

\fig{dec28_state_vectors}{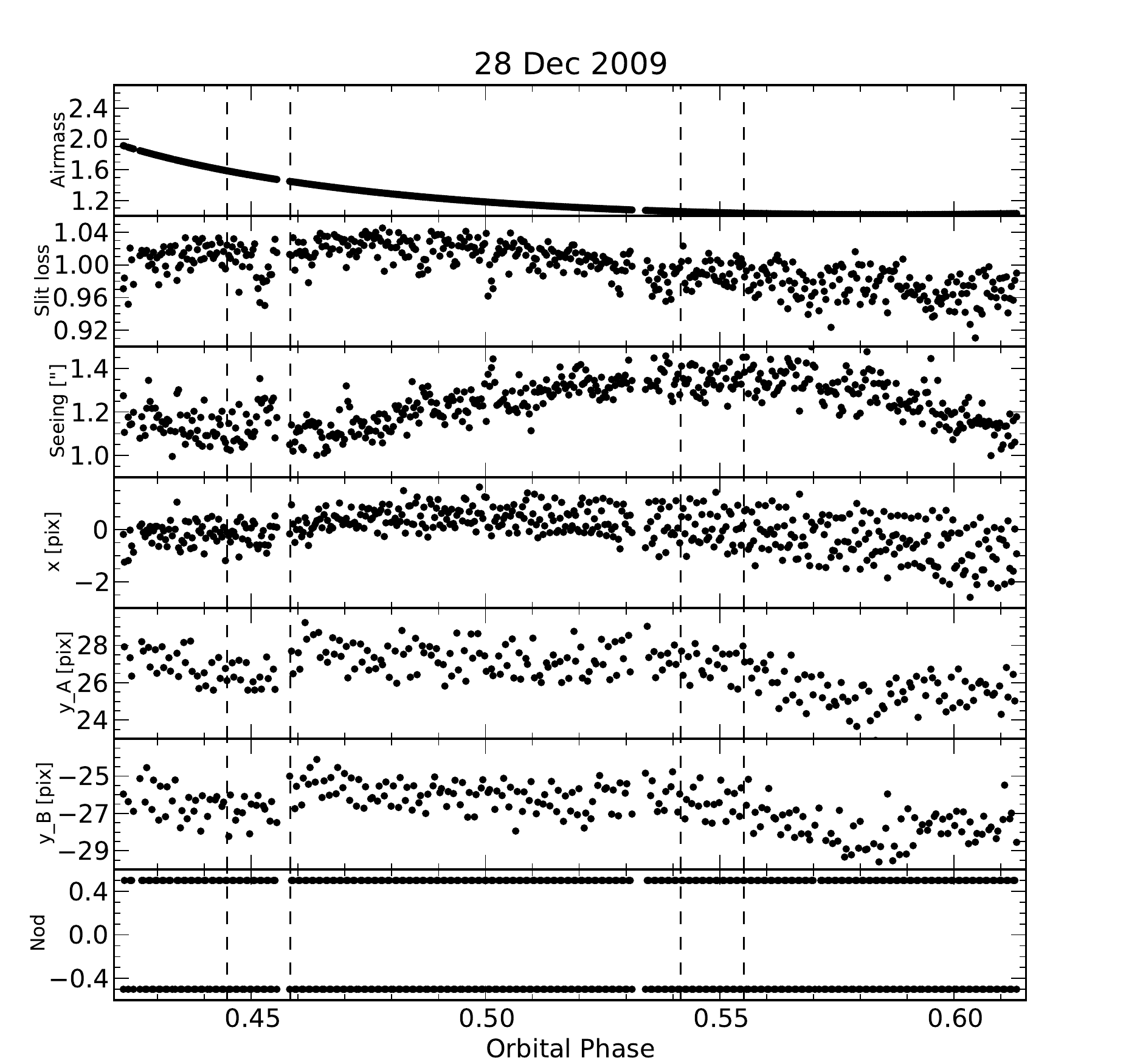}{width=3.5in}{}{The
  observable quantities (described in Sec.~\ref{sec:reduction})
  measured during the course of our observations on 28~Dec~2009.  As
  described in the text, we ultimately detrend our observations with a
  combination the airmass reported by the telescope control system and
  the nod position vector.  As noted in the text we measure the seeing
  FWHM and $y$ position as a function of wavelength, but here we plot
  only the approximate K-band values of these quantities. The dashed
  lines indicate the four points of contact of the eclipse as
  calculated from the ephemeris of \cite{hebb:2009}.}

\fig{dec30_state_vectors}{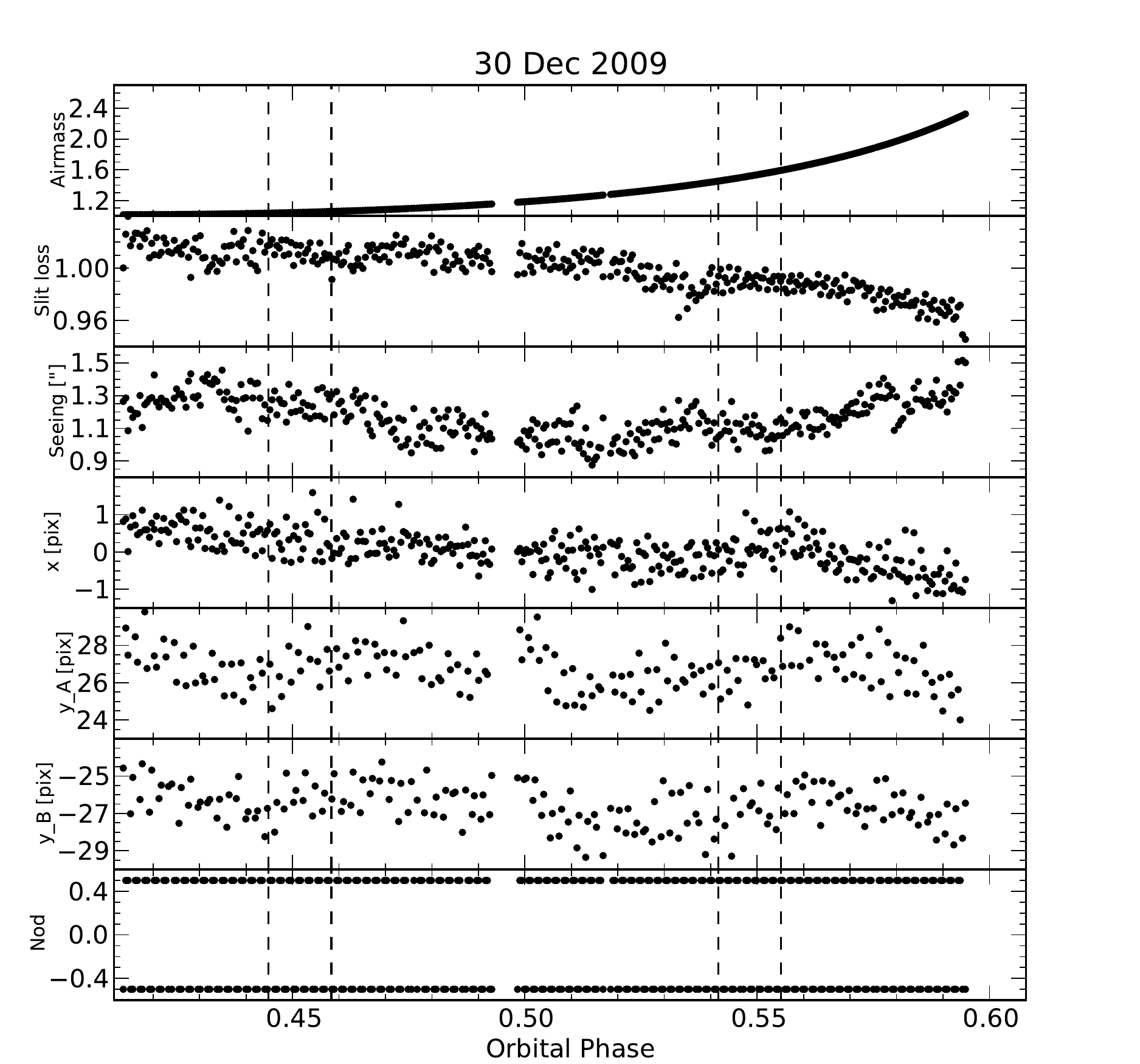}{width=3.5in}{}{Same
  as Fig.~\ref{fig:dec28_state_vectors}, but for the night of
  30~Dec~2009.}

SpeX is a large instrument and is mounted at the IRTF's Cassegrain
focus, where its spectra can exhibit several pixels of flexure due to
changing gravity vector; similarly, atmospheric dispersion
\citep{filippenko:1982} introduces many pixels of motion at shorter
wavelengths (because we keep the star in the slit by guiding at
K~band).  Apparent spectrophotometric variations can be induced by
such instrumental changes \citep[e.g.,][Paper I]{knutson:2007a}.  We
measure the motion of the spectral profiles in the raw frames
perpendicular to ($x$) and parallel to ($y$) the long axis of the
spectrograph slit as follows.  We compute the $x$ motion of the star
on the slit while aligning the spectra to a common reference frame as
described in Sec.~\ref{sec:reduction} above.  For $y$ we fit Gaussian
profiles to the raw spectral traces, then fit a low-order polynomial
to the measured positions in each frame. The $x$ and $y$ motions are
typically 2-4 pixels in K band and are plotted for both nights in
Figs.~\ref{fig:dec28_state_vectors} and~\ref{fig:dec30_state_vectors}.
An independent method to measure the $x$ and $y$ motions would be to
use images recorded by SpeX's slit-viewing camera: since the slit is
slightly reflective one would then be able to measure directly the
star's position on the slit at the guiding wavelength.  We recommend
observers investigate this approach in the future.

We measure the full-width at half maximum (FWHM) of the spectral
profiles during the spectral fitting and tracing described above.
Again, we fit a low-order polynomial to the measured values to
smoothly interpolate the compute values.  The value we measure
\citep[which does not scale as $\lambda^{-1/5}$ as would be expected
from atmospheric Kolmogorov turbulence;][]{quirrenbach:2000} presumably
depends on a combination of atmospheric conditions, instrumental
focus, and pointing jitter during an exposure, but we hereafter refer
to it merely as seeing.

Previous studies \citep[][Paper~I]{deming:2005} report that an
empirical measure of atmospheric absorption is preferable to the
calculated airmass value when accounting for telluric extinction.  We
measured the flux in a number of telluric absorption lines for the
species CO$_2$, \methane , and H$_2$O in a manner similar to that in
Paper~I.  However, in our empirical airmass terms we
still see substantial contamination from both slit losses and A/B
nodding, and so in our final analysis we use the airmass values
reported by the telescope control system and plotted in
Figs.~\ref{fig:dec28_state_vectors} and~\ref{fig:dec30_state_vectors}.

\subsection{Slit Loss Effects }
\label{sec:sysrem} Absolute spectrophotometry is difficult with narrow
slits because guiding errors, seeing variations, and (when the slit is
not aligned to the parallactic angle) atmospheric dispersion, all
result in a time-varying amount of starlight coupled into the
spectrograph slit \citep[e.g.,][Paper I]{knutson:2007a}. After
extracting the spectra, our next step is to remove the large-scale
flux variations present in the data.  

As described in Sec.~\ref{sec:atmodisp} we try to empirically
calibrate the amount of light entering the spectrograph slit.  Despite
considerable effort, we are only able to qualitatively match the
variability in our observations. This could be because the PSF
morphology (and especially the wavelength-dependent flux ratio between
the core and wings) cannot be accurately modeled using a simple
Gaussian function (perhaps due to alignment errors within SpeX and/or
guiding errors), because our implementation of the simplified
formulation of \cite{green:1985} does not reflect reality with
sufficient fidelity, or because variations due to telluric sources
overwhelm those due to instrumental effects. An independent test could
be performed in future efforts by recording images from the
slit-viewing camera and directly measuring the light not entering the
slit, the shape of the PSF, and its position.

Instead, following Paper~I we divide the flux in every wavelength
channel by a wavelength-independent slit loss time series.  This step
removes the absolute eclipse depth (the mean depth over the slit loss
wavelength range) from all spectral channels, but the overall shape of
the emission spectrum should remain the same. However, the quality of
this correction will degrade rapidly at shorter wavelengths because
air's refractive index increases rapidly at shorter wavelength.
Especially with a narrow slit (as during our 28~Dec observations) or
at high airmass (as on 30~Dec), this can cause a greater proportion of
the short-wavelength flux to fall outside the slit. Nonetheless, we
are unwilling to venture beyond removal of this simple achromatic
trend, given our inability to accurately model the chromatic slit loss
component.

Dividing the data by this time series substantially reduces the
variability in regions clear of telluric absorption, as shown in
Figs.~\ref{fig:corrdata}, \ref{fig:dec28_phot}, and
\ref{fig:dec30_phot}.  Note however that some correlated variability
remains even after this correction step, as seen for example near
orbital phase~0.45 on 28~Dec (Figs.~\ref{fig:corrdata}
and~\ref{fig:dec28_phot}). These residual variations are
wavelength-dependent, and support our conclusion that chromatic slit
losses are affecting our data. Wider slits should reduce this effect,
and indeed such chromatic residuals are reduced by a factor of $\sim2$
on 30~Dec (see Fig.~\ref{fig:corrdata}), when we used the wider slit.

\figtwocol{corrdata}{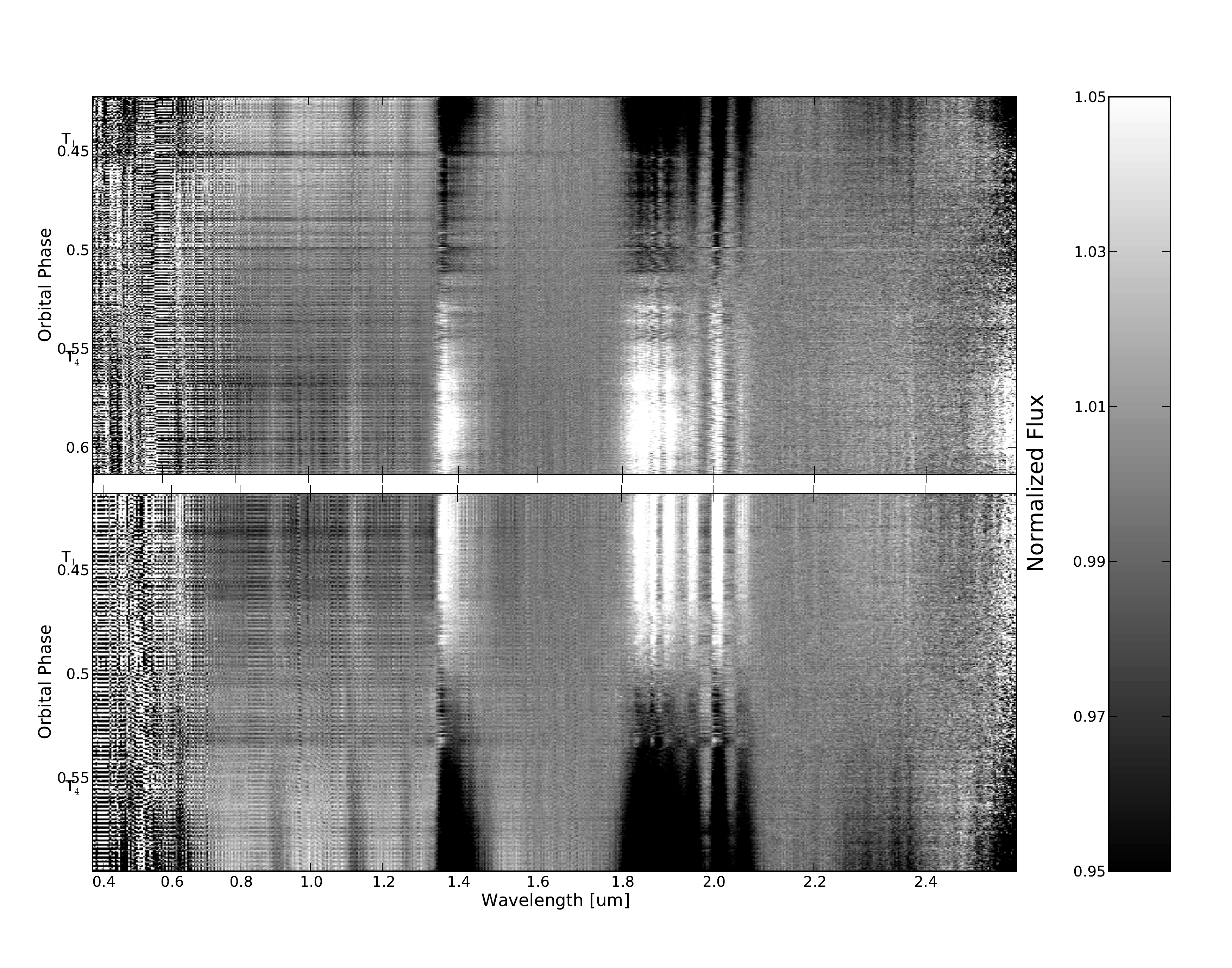}{width=7in}{}{Spectrophotometric data for
  the nights of 28~Dec ({\em top}) and 30~Dec ({\em bottom}) after
  dividing all wavelength channels by the achromatic slit loss time
  series and normalized by the median flux in each wavelength channel.
  Still no eclipse is visible because dividing by the achromatic slit
  loss term has removed the mean eclipse signal from all wavelength
  channels, but variations have been strongly suppressed.  The first
  ($T_1$) and fourth ($T_4$) points of contact of the eclipse are
  noted, as calculated from the ephemeris of \cite{hebb:2009}.}

\fig{dec28_phot}{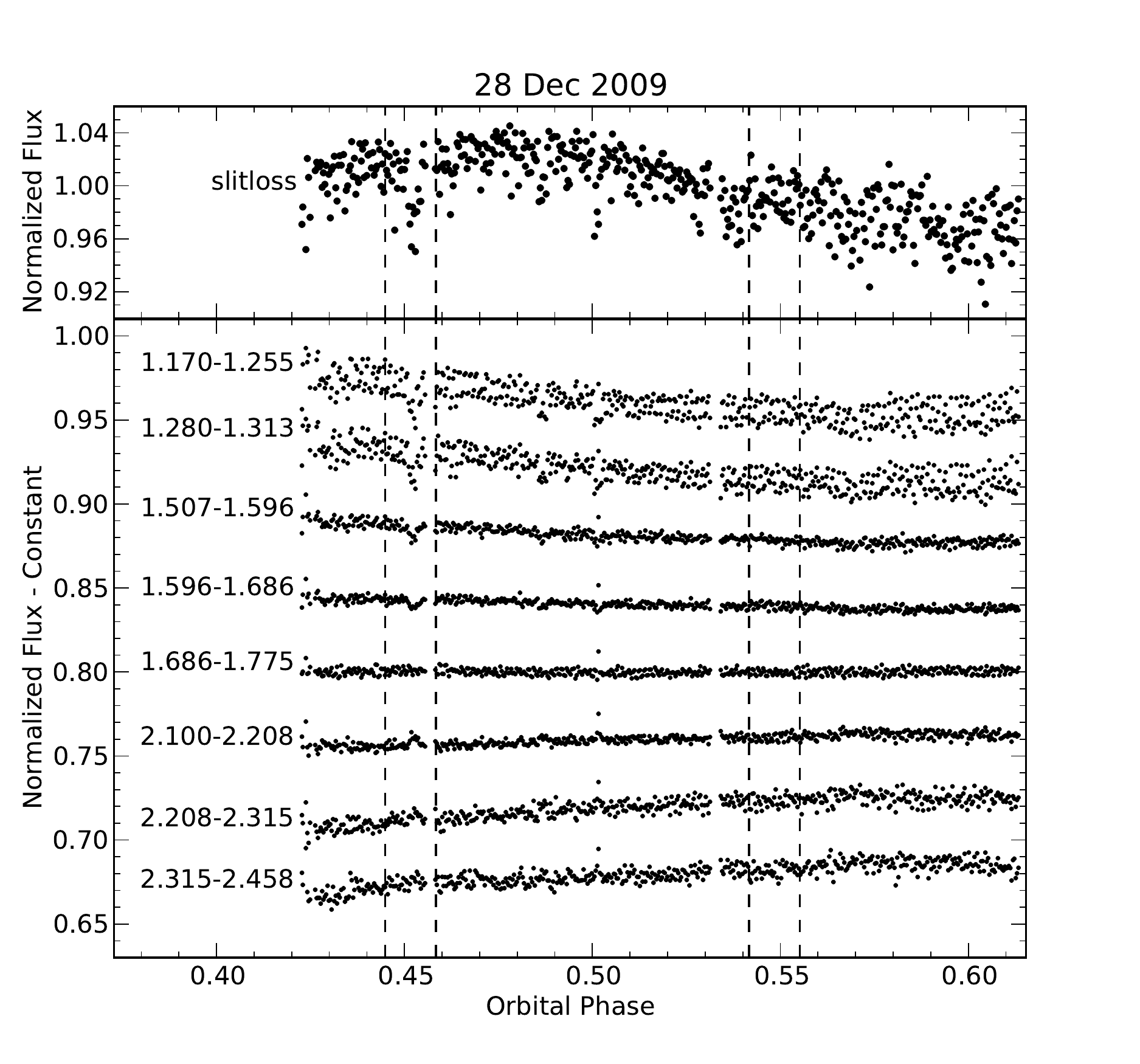}{width=3.5in}{}{Several representative
  spectrophotometric time series for 28~Dec.  The top panel shows the
  relative flux coupled into the spectrograph slit, as measured in
  regions free of deep telluric absorption lines; telluric continuum
  absorption, seeing variations, and guiding errors combine to produce
  large variations, wholly masking the $\lesssim0.3$\% eclipse
  signature.  The bottom panel shows time series for several different
  wavelength ranges, after removal of the common mode slit loss term
  and binned over the wavelength range listed (in \micron).  The
  eclipse is still not visible because dividing out the common-mode
  slit loss term removes the mean eclipse signal from all the
  data. Dashed lines are as in the previous figures.}

\fig{dec30_phot}{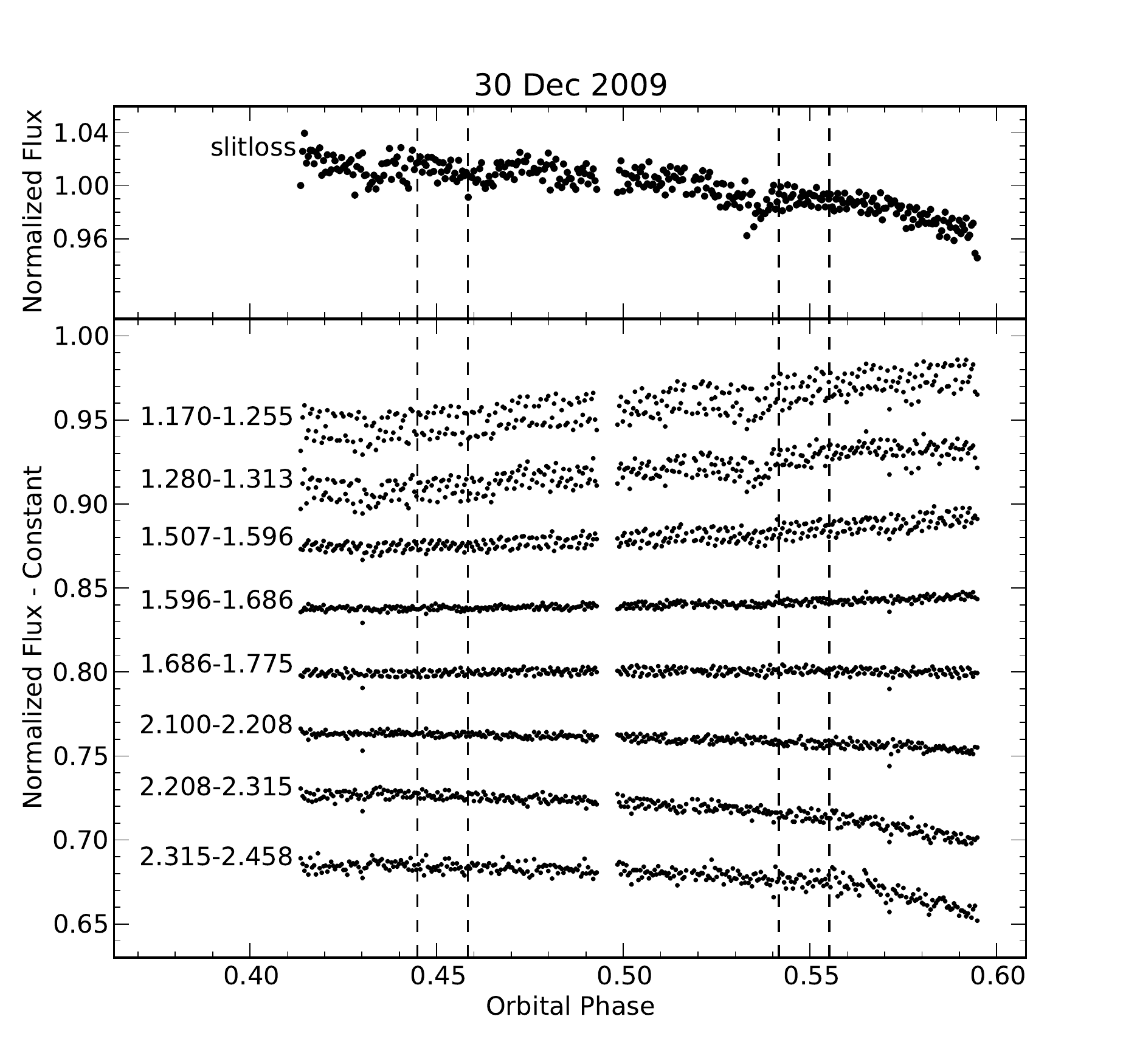}{width=3.5in}{}{Same as
  Fig.~\ref{fig:dec28_phot}, but for the night of 30~Dec. Note that
  these data are less noisy than those shown in the previous figure,
  probably because of the different slit sizes used.}

Our simple correction reveals a residual sawtooth-like pattern in the
photometry in phase with the A/B nodding and especially prominent at
shorter wavelengths ($<1.4$\,\micron), as seen in
Figs.~\ref{fig:corrdata}, \ref{fig:dec28_phot}, and
\ref{fig:dec30_phot}.  The sawtooth has been previously noted with
SpeX in echelle mode \citep{swain:2010} and presumably results from an
imperfect flat-field correction of the differential sensitivity
between the two nod positions on the detector.  We fit the data at
both positions simultaneously by including a vector equal to $0.5$ at
the A nods and $-0.5$ at the B nods in our set of potential
systematic-inducing observables (as described in
Sec.~\ref{sec:detection} below).  That the sawtooth is stronger at
shorter wavelengths may indicate that the fidelity of the SpeX
internal flat fields is wavelength-dependent. Since in any case the
eclipse signal is stronger at longer wavelengths \citep{croll:2011a},
and because the shorter-wavelength regions experience larger motions
on the detector due to atmospheric refraction and larger systematic
biases due to chromatic slit losses (described in
Sec.~\ref{sec:atmodisp}), we ultimately discard the
shortest-wavelength data.

\section{Searching for the Eclipse Spectrum}
\label{sec:detection} 
\subsection{Fitting to the Data}
As noted previously, without external calibration we cannot accurately
recover the absolute eclipse depth from the telluric-contaminated
spectrophotometry.  Instead, we self-calibrate as described in
Sec.~\ref{sec:sysrem} above by dividing out a common time series,
thereby largely removing systematic effects (such as variable slit
loss); information about the absolute eclipse depth is lost, but the
shape of the spectrum is largely unchanged (note however that
systematic effects remain that will influence the extracted planetary
spectrum; we quantify these effects in Sec.~\ref{sec:sys} below).  We
are then better able to look for the eclipse signature as a
differential effect while relying on the precise NIR photometric
eclipse depths \citep{croll:2011a} to place our measurements on an
absolute scale.  However, even after removing the common-mode time
series the eclipse signal is still masked by the photometric sawtooth,
airmass dependencies, and general photometric noise.

We cannot use cross-correlation techniques
\citep[][Paper~I]{deming:2005, snellen:2010} in this analysis because
of our low resolution. We investigated the use of the Fourier-based
self-coherence spectrum technique \citep{swain:2010} but did not find
it to remove correlated variability or to otherwise improve the
quality of our data. Instead, we follow Paper~I and search for
differential eclipse signatures in our data by fitting a model that
includes telluric, systematic, and eclipse effects to the slit
loss-corrected time series in each wavelength channel; this approach
also has the advantage of allowing an estimate of the covariances of
the various determined parameters.

We fit each spectral time series (i.e., the flux in each wavelength
bin) with the following relation, representing an eclipse light curve
affected by systematic and telluric effects:
\begin{equation}
  F^\lambda_i = f^\lambda_0 \left( e^{b^\lambda {a_i}} \right) \left( 1 + d^\lambda \ell_i \right) 
  \left( 1 + \sum_{j=1}^J c^\lambda_{j}v_{ij} \right)  
\label{eq:fluxeqn}
\end{equation}  
The symbols are: $F^\lambda_i$, the slit loss-corrected flux measured
at timestep $i$ in wavelength bin $\lambda$; $f^\lambda_0$, the total
(star plus planet day side) flux that would be measured above the
Earth's atmosphere; $a_i$, the airmass, which is modulated by the
coefficient $b^\lambda$, an airmass-like extinction coefficient in
which the airmass is proportional to the log of observed flux;
$\ell_i$, the flux in an eclipse light curve scaled to equal zero out
of eclipse and -1 inside eclipse; $d^\lambda$, a scale parameter equal
to the relative depth of eclipse; $v_{ij}$, the $J$ state vectors
(e.g., nod position, $x_i$ or $y_i$) expected to have a small,
linearly perturbative effect on the instrumental sensitivity; and
$c^\lambda_j$, the coefficients for each state vector.  To account for
and remove the effect of any slow drifts we also tried including
low-order Chebychev polynomials in orbital phase in the set of state
vectors, but these did not improve our results. We thus obtain the set
of coefficients $(f_0^\lambda, d^\lambda, c_j^\lambda )$ from our full
set of observations; the $d^\lambda$ represent our measured emission
spectrum.

To fix the parameters of our model eclipse light curve we compared the
orbital ephemerides from several different sources \citep{hebb:2009,
  campo:2011, croll:2011a, chan:2011} and found them all to be
consistent to within 1-2 minutes at our observational epoch, an
uncertainty insignificant given the noise in our data and our sampling
rates.  We therefore use the parameters from \cite{hebb:2009}, which
we compute using our Python implementation\footnote{Available from the
  primary author's website; currently
  \protect{\url{http://astro.ucla.edu/~ianc/python/transit.html}}} of
the uniform-disk formulae of \cite{mandel:2002}.

\subsection{Choice of Model}
As in Paper~I, we fit the data sets using many different combinations
of state vectors and slit loss time series and use the BIC to choose
which of these many models best fit our data.  Calculating the BIC for
each set of parameters involves computing $\chi^2$ for each time
series, which in turn requires us to assign uncertainties to each data
point.  We estimate the uncertainties as follows.  We initially
compute unweighted fits of Eq.~\ref{eq:fluxeqn} to the data using a
multivariate minimization provided in the SciPy\footnote{Available at
  \url{http://www.scipy.org/}.} software distribution (the function
\texttt{optimize.leastsq}).  Decorrelating using only the A/B nod
position and airmass calculated from the telescope's zenith angle, we
fit and compute the residuals for each time series.  We scale the
uncertainties in each time series such that the $\chi^2$ in each
wavelength channel equals unity.  For each combination of state
vectors we then compute another, weighted, fit and its associated
$\chi^2$ and BIC.  Although this method of estimating uncertainties
likely underestimates absolute parameter uncertainties \citep[][and
see Sec.~\ref{sec:errors} below]{andrae:2010}, we feel it still allows
us to compute useful qualitative estimates of the relative merit of
various models.

Our modeling approach is most successful in spectral regions largely
clear of telluric absorption, which suggests telluric absorbers may be
one of the primary factors limiting our analysis (as confirmed in
Sec.~\ref{sec:telluric} and~\ref{sec:allsys}).  When restricting our
analysis to the BIC values computed in regions largely clear of strong
telluric effects ($1.52-1.72$\,\micron\ and $2.08-2.34$\,\micron), the
instrumental models which give the lowest BIC for our data use a slit
loss term computed using telluric-free spectral regions in the H band,
the airmass values reported by the telescope control system, and two
state vectors: the A/B nod position and an airmass-corrected,
mean-subtracted copy of the slit loss term.  The BIC values do not
change significantly when we use slightly different wavelength ranges.

Although including these two decorrelation vectors appears warranted
on statistical grounds, our modeling efforts (discussed in
Sec.~\ref{sec:allsys}) demonstrate that decorrelating against the slit
loss time series in the light curve fits systematically biases the
extracted planetary spectrum.  Because the slit loss effects removed
by including this vector are chromatic, the coefficient associated
with this vector increases at shorter wavelengths.  Since our
achromatic slit loss vector is not wholly orthogonal to the model
eclipse light curve, as the slit loss vector's amplitude increases the
eclipse depth tries to compensate, and the extracted spectrum is
corrupted. Our modeling of the 30~Dec observations (when the 3.0''
slit was used) indicates that for these data this bias would mainly
affect $\lambda < 1.4\,\micron$, but the bias is stronger for the
28~Dec data (when the 1.6'' slit was used) and significantly affects
the H~band as well.  Thus we again emphasize that similar observations
in the future should use as large a slit as possible, and should guide
at the parallactic angle, in order to mitigate the biases introduced
by chromatic slit loss. For these reasons we include only the A/B nod
vector in our list of decorrelation vectors

\subsection{Estimating Coefficient Uncertainties}
\label{sec:errors}


We assess the statistical uncertainties on the computed planetary
spectra using several techniques.  First, we fit to the data in each
of the 564 wavelength channels as described above and compute the mean
and standard deviation of the mean (SDOM) of the parameters in
wavelength bins of specified width. The SDOM provides a measure of the
purely statistical variations present in the planetary spectra.

After summing the data into wavelength channels 25~nm wide (to ease
the computational burden) we run both Markov Chain Monte Carlo (MCMC)
and prayer bead \citep[or residual permutation;][]{gillon:2007}
analyses for each wavelength-binned time series.  Since MCMC requires
an estimate of the measurement uncertainties, we follow our earlier
approach of setting the uncertainties in each wavelength channel such
that the resultant $\chi^2$ value equals unity. The residual
permutation method fits multiple synthetic data sets constructed from
the best-fit model and permutations of the residuals to that fit, and
it is similar to bootstrapping but has the advantage of preserving
correlated noise.

The posterior distributions of eclipse depth that result from the MCMC
analysis are all much narrower than the uncertainties estimated from
both the SDOM and from the prayer bead analysis. This suggests that
artificially requiring that $\chi^2$ equal unity has led to
underestimated parameter uncertainties \citep[cf.][]{andrae:2010}.
The prayer bead and SDOM uncertainties are comparable in magnitude,
and to be conservative we use the larger of these two uncertainties in
each wavelength bin as our statistical uncertainty.

Because we expect systematic uncertainties to play a large role in our
data, in the following section we now pause to examine possible
sources of bias and their impact on our planetary spectra.

\section{Systematic Errors in High-Precision Single-Slit Spectroscopy}
\label{sec:sys}
Our analysis is hampered by systematic biases arising from several
sources.  We discuss telluric contamination arising from variable
transmittance and/or radiance (which affects only certain wavelength
ranges) in Sec.~\ref{sec:telluric}.  In Sec.~\ref{sec:atmodisp} we
discuss chromatic slit losses, which result from wavelength-dependent
seeing and atmospheric dispersion; this introduces a smoothly varying
bias across the entire spectrum, increasing in severity toward shorter
wavelengths.  Then we combine these effects in Sec.~\ref{sec:allsys}
and use all available information to simulate our
observations. Applying our standard reduction to these simulations
demonstrates that we can hope to successfully recover a planetary
signal within certain well-defined spectral regions.

\subsection{Telluric Contamination}
\label{sec:telluric}
Increased levels of precipitable water vapor (PWV) lead to increased
telluric emittance and decreased transmittance.  If unaccounted for,
such variations can mimic and/or contaminate the desired eclipse
spectrum \citep[][but see also
\citeauthor{waldmann:2011}~\citeyear{waldmann:2011}]{mandell:2011}.
The claim of a strong ground-based L band detection of HD~189733b in
eclipse \citep{swain:2010} was challenged partially by an appeal to
changes in telluric water content \citep{mandell:2011}, so we
investigate these effects in our observations.

As can be seen in Fig.~\ref{fig:dec28_fitcoefs}, the 28~Dec eclipse
spectrum is strongly biased toward larger eclipse depths in regions of
greater telluric absorption. This does not seem to be the case for the
30~Dec results (cf. Fig.~\ref{fig:dec30_fitcoefs}), in which we see
variability (but no net deflection of the spectrum) in regions of high
telluric absorption.  This behavior suggests that our data are
compromised by telluric effects in these wavelength ranges, and the
regions of greatest spectral deflection suggest telluric water vapor
is the prime culprit.

\fig{dec28_fitcoefs}{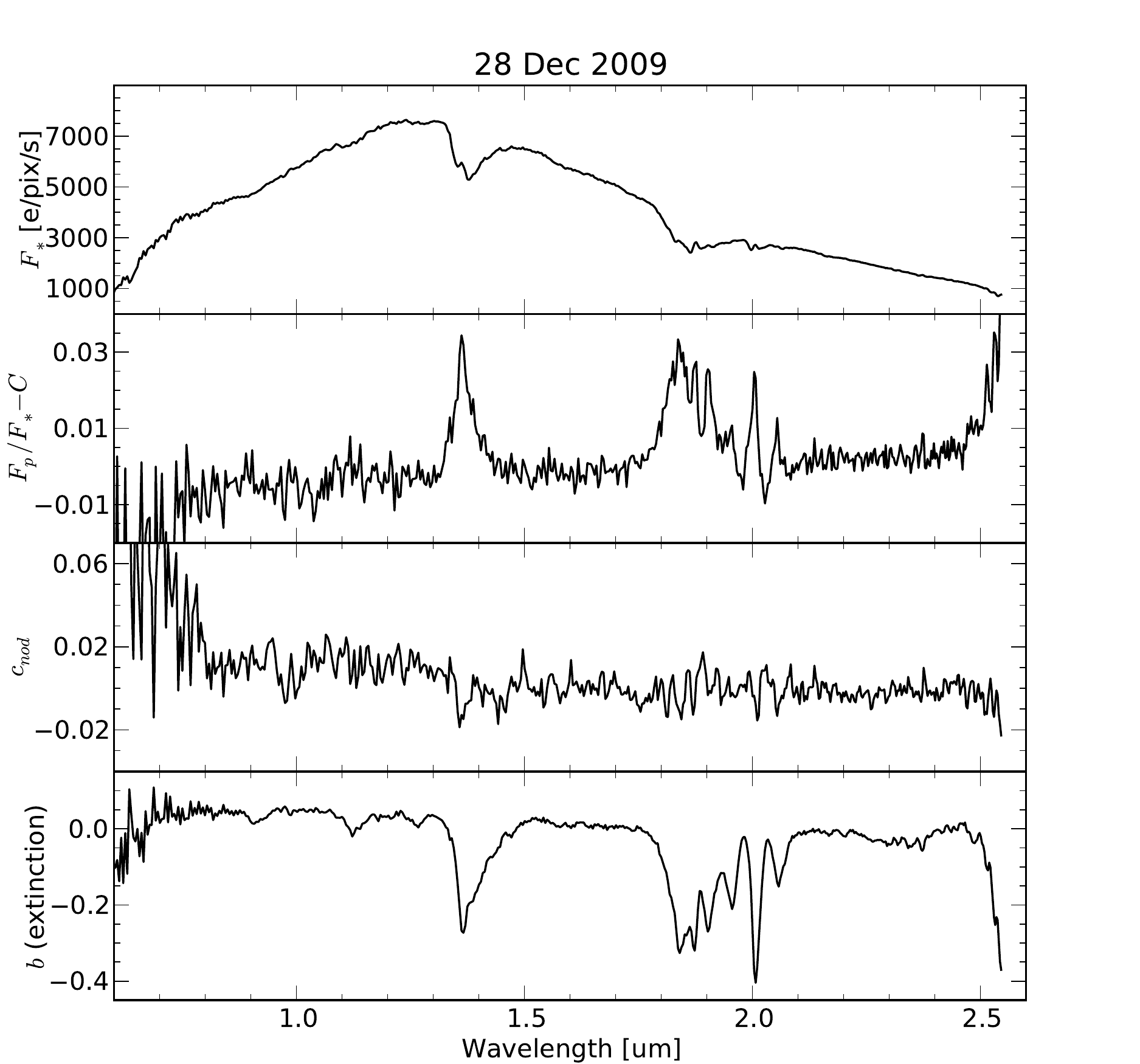}{width=3.5in}{}{Best-fit
  coefficients from fitting Eq.~\ref{eq:fluxeqn} to the slit
  loss-corrected 28~Dec observations  shown in
  Fig.~\ref{fig:corrdata}. From top to bottom: stellar flux, eclipse
  depth, A/B nod sensitivity coefficient, and telluric extinction
  coefficient. Refer to Sec.~\ref{sec:detection} for a description of
  the fitting process. }

\fig{dec30_fitcoefs}{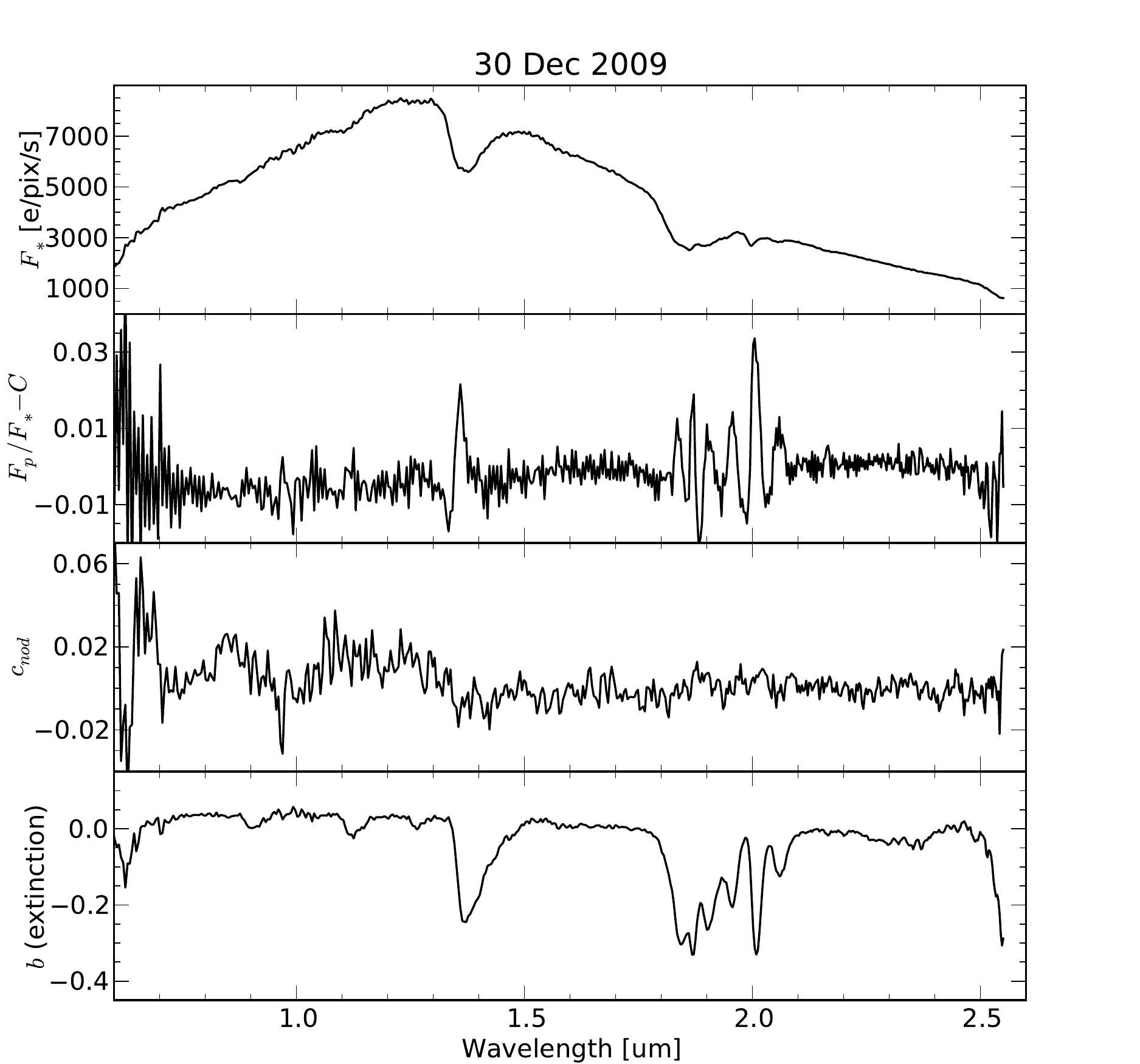}{width=3.5in}{}{Same as
  Fig.~\ref{fig:dec28_fitcoefs}, but for the night of 30~Dec.}

Telluric water content is measured on Mauna Kea by the 350\,\micron\
tipping photometer at the Caltech Submillimeter
Observatory\footnote{Data taken from
  \url{http://ulu.submm.caltech.edu/csotau/2tau.pl} }.  We convert its
350\,\micron\ opacity measurements to PWV using the relation from
\cite{smith:2001}:
\begin{equation}
PWV = 20 ( \tau_{350} / 23 - 0.016) \textrm{~mm}
\end{equation}
The PWV values for the two nights we observed are plotted in
Fig.~\ref{fig:tau}. Although the PWV along the telescope's line of
sight will scale with airmass, because our fitting approach removes
airmass-correlated trends we consider only the water burden at
zenith. On 28~Dec the mean PWV values in and out of eclipse were 0.64
and 0.60~mm, respectively; on 30~Dec these values were 0.68 and
0.70~mm, respectively.

\fig{tau}{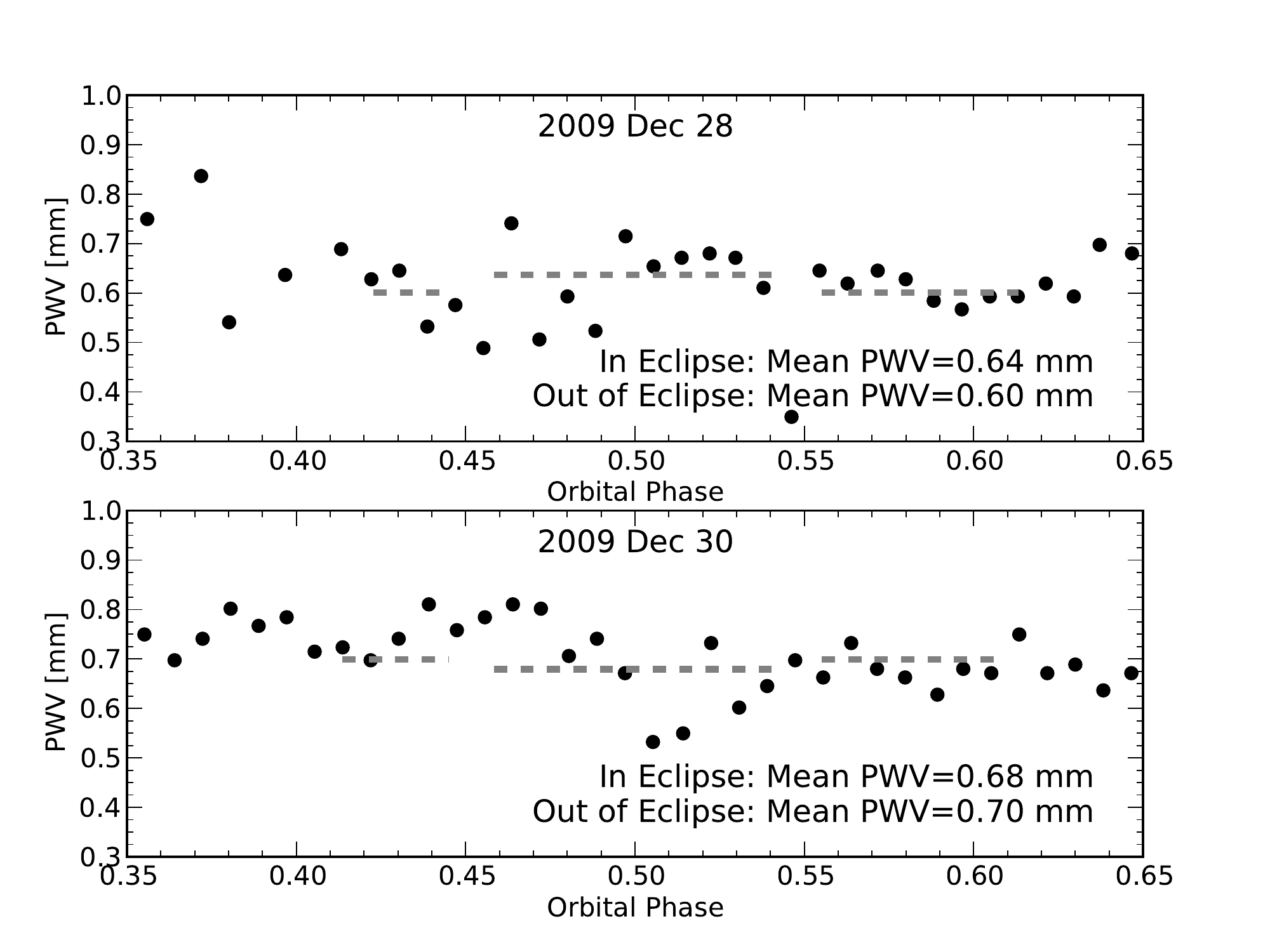}{width=3.5in}{}{Telluric water content during
  our observations, as measured by the 350\,\micron\ tipping
  photometer at the Caltech Submillimeter Observatory.  The dashed
  lines represent the mean PWV values in and out of eclipse on each of
  the two nights, and also indicate the start and end of each nights'
  observations.}

We used two independent telluric modeling codes, ATRAN
\citep{lord:1992} and LBLRTM\footnote{Run using MATLAB scripts made
  publicly available by D.~Feldman and available at
  \url{http://www.mathworks.com/matlabcentral/fileexchange/6461-lblrtm-wrapper-version-0-2}}
\citep[Version 12.0;][]{clough:2005}, to generate NIR telluric spectra
for the in- and out-of-eclipse PWV values; all spectra were computed
using an airmass of unity.  ATRAN simulates atmospheric transmission
only, while LBLRTM simulates both transmission and emission.  The
apparent eclipse signal induced by transmission changes is
$\Delta\textrm{Tran} = (t_{out} - t_{in})/t_{out}$, where $t_{in}$ and
$t_{out}$ are the in- and out-of-eclipse transmission spectra; the
radiance-induced signal is $\Delta\textrm{Rad} = \Omega(s_{out} -
s_{in})/(s_* + \Omega s_{out})$, where $s_{in}$ and $s_{out}$ are the
sky radiance spectra in and out of eclipse, $s_*$ is the incident
stellar flux, and $\Omega$ is the solid angle on the sky of the
effective spectral extraction aperture. We validated our models
against the study of \cite{mandell:2011} and match their results to
within 15\%, which we deem an acceptable match given the large number
of user-specified parameters in such simulations. While we thus
confirm that the $3-3.5$\,\micron\ L band spectrum reported by
\cite{swain:2010} for \hdoneb\ appears similar to the spectrum that
would result from uncorrected variations in telluric water vapor
emission, water vapor radiance effects do not match their spectrum
from $3.5-4$\,\micron , where eclipse depths of 0.5\%\ would be seen;
nor do radiance effects match their K band spectrum.  A complete
explanation of the \cite{swain:2010} results must involve more than
merely telluric effects.

Over our wavelength range we find that telluric thermal radiation is
low enough that $|\Delta\textrm{Rad}| < |\Delta\textrm{Tran}|$ always,
so we neglect radiance effects.  We plot the $\Delta\textrm{Tran}$
signals with the observed eclipse spectra in
Fig.~\ref{fig:pwv_effects}, and the comparison is intriguing.  The
28~Dec eclipse spectrum bears a striking resemblance to our calculated
$\Delta\textrm{Tran}$ spectrum, suggesting these observations are
affected by variations in telluric water vapor transmission at some
wavelengths. However, the 30~Dec observations show only a weak
correlation with the $\Delta\textrm{Tran}$ signal (in the wings of
strong water bands), suggesting that the CSO data allow for only a
crude estimate of the effects of atmospheric water on the extracted
planetary spectrum.

\fig{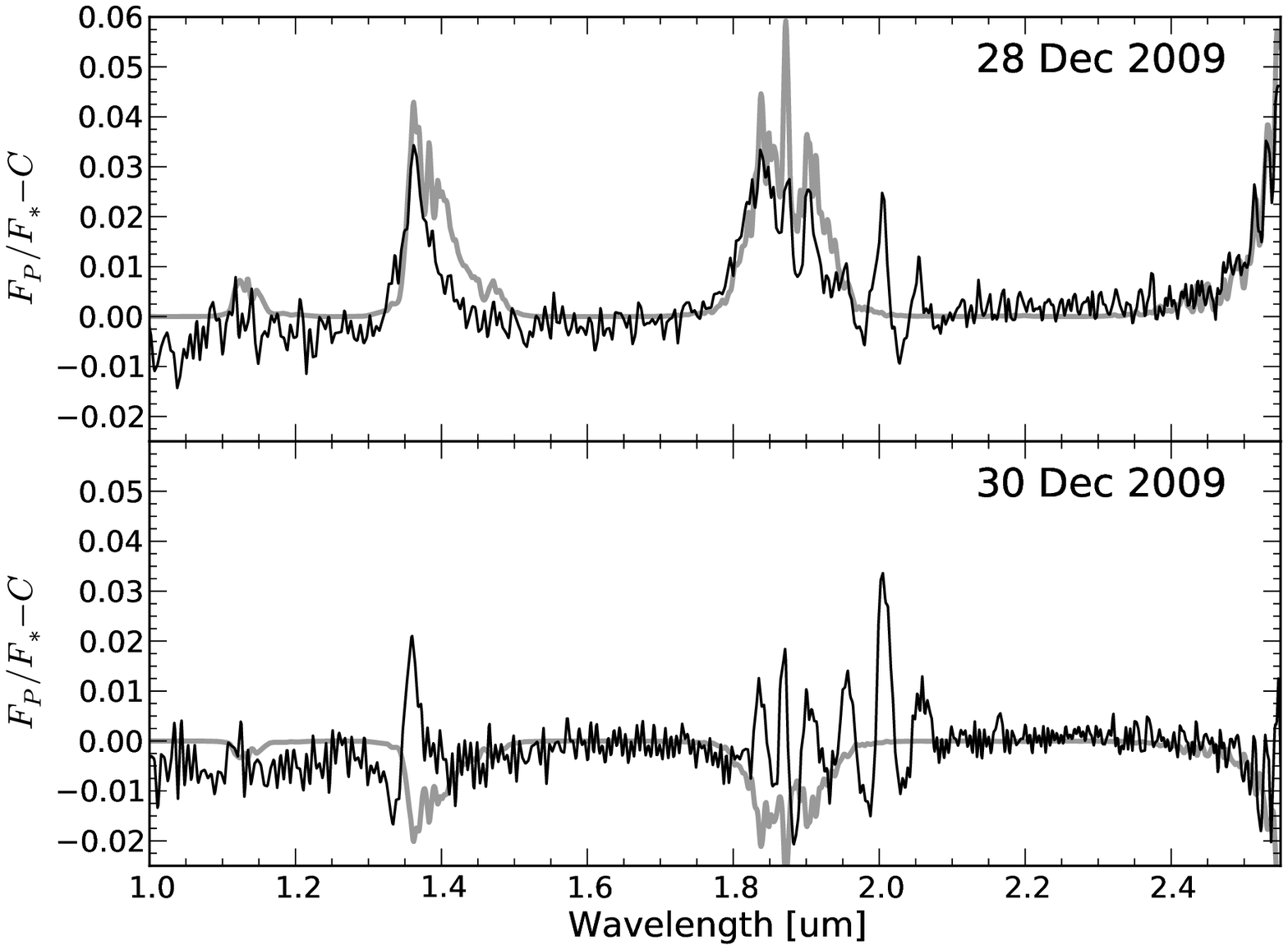}{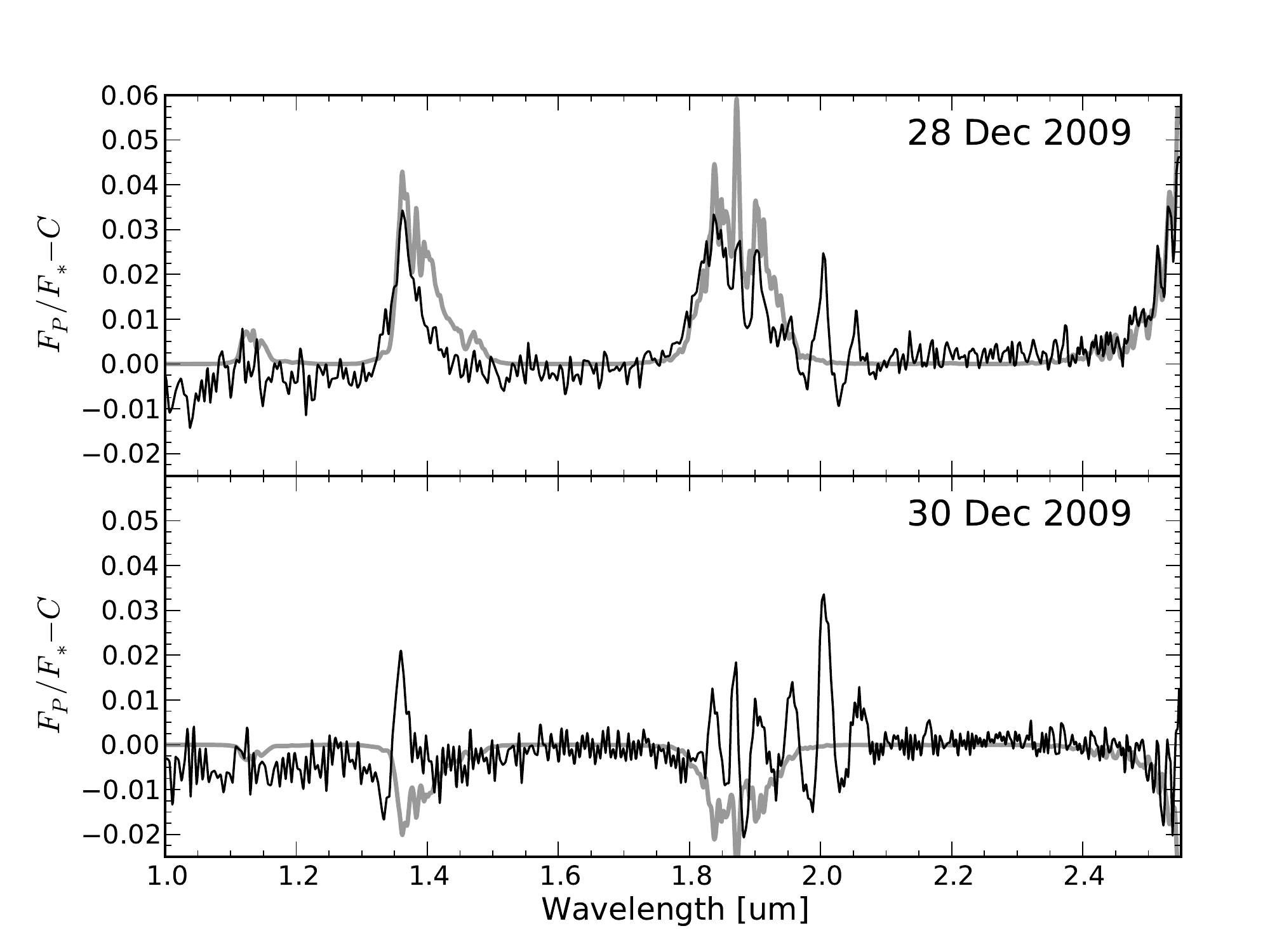}{width=3.5in}{}{The effect of changes
  in telluric water absorption during our observations. The thin black
  line shows the measured residual eclipse spectrum, while the thick
  gray line represents the apparent eclipse signal
  ($\Delta\textrm{Tran}$) that would be inferred from the uncorrected
  changes in telluric PWV shown in Fig.~\ref{fig:tau}. The 28~Dec
  spectrum appears strongly correlated with the $\Delta\textrm{Tran}$
  signal, but the 30~Dec spectrum does not.  Neither
  $\Delta\textrm{Tran}$ spectrum is significant far from telluric
  water absorption lines.}

For both nights, the $\Delta\textrm{Tran}$ spectra do not capture the
large spectral variations in the eclipse spectra from
$2-2.07$\,\micron\ where there are strong telluric \coo\ absorption
bands. We generate several ATRAN atmospheric profiles with varying
concentrations of \coo\ but find that the in- and out-of-eclipse \coo\
concentrations must differ by $>5$~ppm to reproduce the features seen
at these wavelengths.  Such a change would be greater than any
hour-to-hour change recorded at Mauna Loa by the National Oceanic and
Atmospheric Administration Earth System Research Laboratory (NOAA
ESRL) during all of 2009 \citep{thoning:2010}.  Thus the telluric
residuals in this wavelength range, though clearly correlated with the
telluric \coo\ bands, are more likely attributable to the
non-logarithmic relationship between flux and airmass in
near-saturating lines and not to time-variable concentration.

As noted, $\Delta\textrm{Rad}$ is negligible across most of our
passband, reaching $< 2 \times 10^{-4}$ by 2.4\,\micron\ for our PWV
values.  The magnitude of $\Delta\textrm{Tran}$ shown in
Fig.~\ref{fig:pwv_effects} is $<2 \times 10^{-4}$ for our observations
in the wavelength ranges $1-1.1$\,\micron, $1.22-1.30$\,\micron,
$1.52-1.72$\,\micron, and $2.03-2.34$\,\micron. We further exclude the
spectral regions affected by \coo\ ($2.00-2.08$\,\micron).  So long as
we restrict our analysis to these regions we consider it unlikely that
telluric water or \coo\ significantly affect our results on either
night.

Methane is another species whose abundance we are interested in
measuring but whose telluric concentration can vary on short
timescales.  The NOAA ESRL also measures atmospheric \methane\ content
\citep{dlugokencky:2011}, so we examined the hourly logs.  The largest
hour-to-hour change during our observations was $\sim0.5\,\%$, with
typical hourly changes smaller by a factor of several.  We again use
ATRAN \citep{lord:1992} to simulate two atmospheric transmission
spectra with methane amounts varying by 0.5\,\% (PWV was set to 1~mm
and we simulated observations at zenith), and we then calculate
$\Delta\textrm{Tran}$ as before.  At our spectral resolution we find
that $\Delta\textrm{Tran}$ reaches a maximum of about 0.04\,\% near
2.36\,\micron\ and is $<10^{-4}$ outside of $2.23-2.4\,\micron$.  We
include this $\Delta\textrm{Tran}$ spectrum as a wavelength-dependent
systematic uncertainty in our final measurements.

\subsection{Chromatic Slit Losses}
\label{sec:atmodisp}
We quantify the impact of chromatic slit loss on our data by modeling
this effect and then trying to extract spectroscopic information from
the simulation. For this modeling we use an implementation based on
\texttt{lightloss.pro} in the SpeXTool \citep{cushing:2004}
distribution; this in turn is based on the discussion of atmospheric
dispersion in \citeauthor{green:1985} (\citeyear{green:1985}; their
Eq.~4.31).  A crucial factor in these simulations is the refractive
index of air, which we model following \cite{boensch:1998} assuming
air temperature, pressure, and composition that are constant but
otherwise consistent with values typical for Mauna Kea.  We also used
our empirical measurements of the wavelength-dependent seeing FWHM and
the positions of the spectra along the slit.  We cannot measure
atmospheric dispersion perpendicular to the slit's long axis, so we
calculate this wavelength-dependent quantity and then shift it by the
spectral offsets measured in Sec.~\ref{sec:reduction}.

The result is a model of our chromatic slit loss which is based almost
wholly on empirical data. We see some agreement between this model and
our spectrophotometric throughput -- e.g., less flux and chromatic
tilt of the spectrum during brief periods of poor seeing.  Though our
modeling can qualitatively reproduce the types of variations seen, in
detail the data are highly resistant to accurate modeling and we
suspect additional dispersion and/or optical misalignments in SpeX may
be to blame.

We suspect that our modeling is also limited by an imperfect knowledge
of the (variable) instrument point spread function: the slit loss is
most dependent on the distribution of energy along the dispersion
direction, but we can only measure this shape perpendicular to the
dispersion direction.  We see 10\%\ variations in the seeing from one
frame to the next (as measured by the standard deviation of the
frame-to-frame change in seeing FWHM) -- whether this represents our
fundamental measurement precision or the level of fluctuations in the
instrument response, this level of variation prevents accurate and
precise modeling of the chromatic slit loss.

Whatever the cause of the disagreement, our model appears
qualitatively similar to the spectrophotometric variations apparent in
our observations.  We therefore proceed to extract a planetary
spectrum after removing an achromatic slit loss term as described in
Sec.~\ref{sec:sysrem}.  Although we input no planetary signal the
spectrum extracted is nonzero because, in general, the projection of
the achromatic slit loss vector onto the model eclipse light curve is
nonzero.  As the chromatic slit losses become more severe at shorter
wavelengths, so too is the extracted planetary signal progressively
more biased in those same regions.  We then perform a pseudo-bootstrap
analysis of the chromatic slit loss: we re-order the modeled slit
transmission series -- i.e., we move the first frame's modeled slit
transmission to the end of the data set and re-fit, then move the
second frame's transmission to the end, and repeat -- and each time
extract a planetary spectrum.

The variations in the extracted spectrum represent a systematic bias
introduced by our wavelength-dependent slit losses.  As expected
observations taken with a wider slit fare better: for the 30~Dec
observations the apparent variations in planetary emission (as
measured by the standard deviation in each wavelength channel) are low
in the H~and K~bands, reaching $\gtrsim 0.1\,\%$ (the approximate
magnitude of the expected signal) in J band and rising shortward.
However, our model of the 28~Dec observations indicates a
substantially higher level of systematics: still low in the K band
(where atmospheric dispersion is lessened; this is also our guiding
wavelength, so variations are low here) but rising steeply with
decreasing wavelength, reaching $\sim 0.5\,\%$ by the H~band.  We
apply these noise spectra to our final measurement uncertainties to
account for the possibility of systematic bias.

We also find that the induced spectral variations tend to depend more
on changes in seeing than on atmospheric dispersion when using a 3.0''
slit. This result suggests that our 30~Dec observations were not
significantly compromised by our decision to lock down the instrument
rotator.

\subsection{Result of Simulated Observations}
\label{sec:allsys}
For completeness, we also combine our two dominant sources of
systematic uncertainties -- telluric absorption and chromatic slit
loss effects -- in a comprehensive model of our observations, using
all empirical data available to us.  We use a stellar template for the
star \citep{castelli:2004} and inject a model planetary spectrum
\citep[][the purple curve in their Fig.~1, in which temperature
decreases monotonically with decreasing pressure and with the largest
predicted 2.36\,\micron\ \methane\ bandhead]{madhusudhan:2011} and
modulated by an analytical eclipse light curve \citep{hebb:2009}.  For
each frame we simulate the telluric transmission for each observation
with LBLRTM \citep{clough:2005}, using the appropriate zenith angle
and atmospheric water content (determined by interpolating the CSO
observations in Sec.~\ref{sec:telluric} to the time of the
observation).  We model the chromatic slit loss as described in the
previous section and do not introduce any measurement noise into these
simulated observations; this is because our goal is only to
investigate the systematic biases the aforementioned effects have on
our spectral extraction procedures.  We also assume the detector
response and instrumental throughput (excluding slit losses) are
constant in time and wavelength.  Any temporal variations in detector
sensitivity will manifest themselves as increased scatter in the
residuals and thus propagate to larger uncertainties in the
prayer-bead analysis.

After generating these simulated spectra, we then send them through
the analysis pipeline described in Sec.~\ref{sec:detection}. We plot
the extracted planetary spectra in Fig.~\ref{fig:simulation}.  As
expected, our analysis performs poorly in regions of strong telluric
absorption due to a combination of changing abundances and the more
complicated behavior of partially saturated absorption lines; the
telluric-induced errors are qualitatively similar to those seen in our
simpler analysis of Sec.~\ref{sec:telluric}, confirming our decision
to avoid these wavelengths.

\fig{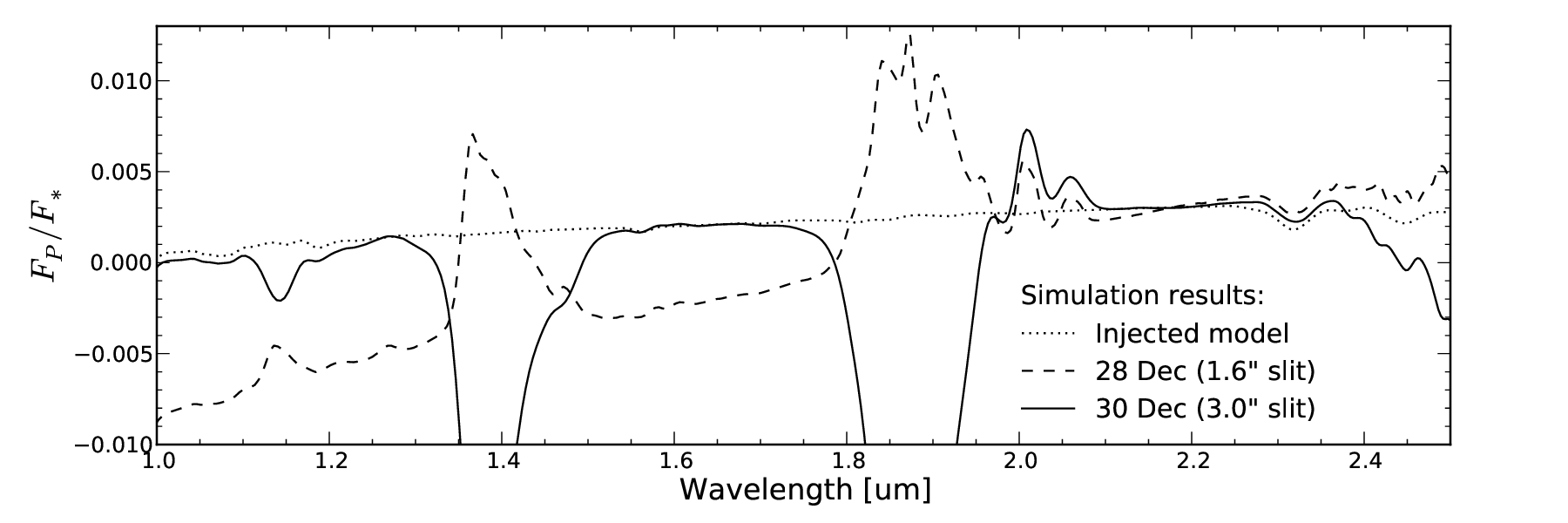}{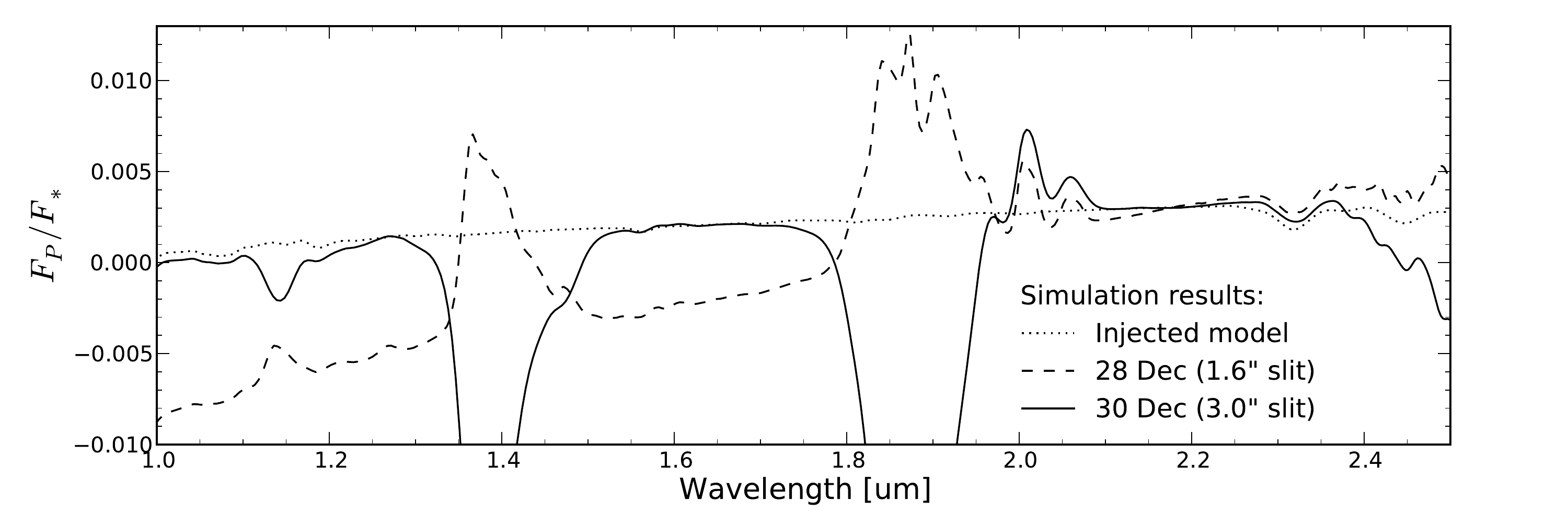}{width=3.5in}{}{Planetary spectra
  extracted from our simulated observations; the difference between
  these and the injected model \citep[from][]{madhusudhan:2011}
  demonstrates the systematic biases present in our data.  Changes in
  telluric water content introduce biases in particular spectral
  regions, while chromatic slit losses introduce gradients across all
  wavelengths, especially with a narrow slit and/or at short
  wavelengths. See Sec.~\ref{sec:allsys} for a complete description.}

Fig.~\ref{fig:simulation} also demonstrates the large systematic bias
introduced by chromatic slit losses.  The effect is especially
pronounced at short wavelengths and, in the case of the narrower
(1.6'') slit, the bias is so large as to prevent this data set from
setting any useful constraints on \wtb's emission.  This finding
agrees with our estimate of the systematic uncertainties induced by
chromatic slit losses in the previous section.

Having developed at least a rough understanding of our data's expected
biases, we are now well-equipped to interpret the results of our
spectroscopic analysis.

\section{Results: Thermal Emission from WASP-12b}
\label{sec:results}
\subsection{Initial Presentation of Results}
The parameters $d^\lambda$ that result from the fitting process
represent a modified emission spectrum of \wtb , specifically
$d^\lambda = F_P^\lambda/F_*^\lambda - C$.  The constant $C$ results
from our correction for common-mode photometric variations, and we set
it using photometric eclipse measurements as described in
Sec.~\ref{sec:results}.  The individual channel (i.e., unbinned)
best-fit coefficients for 28~Dec and 30~Dec are plotted in
Figs.~\ref{fig:dec28_fitcoefs} and~\ref{fig:dec30_fitcoefs},
respectively, and we show the fit residuals in Fig.~\ref{fig:resdata}.
We plot the binned eclipse spectra and their uncertainties (the
quadrature sum of statistical and systematic errors) in
Fig.~\ref{fig:spectrum}.

\figtwocol{resdata}{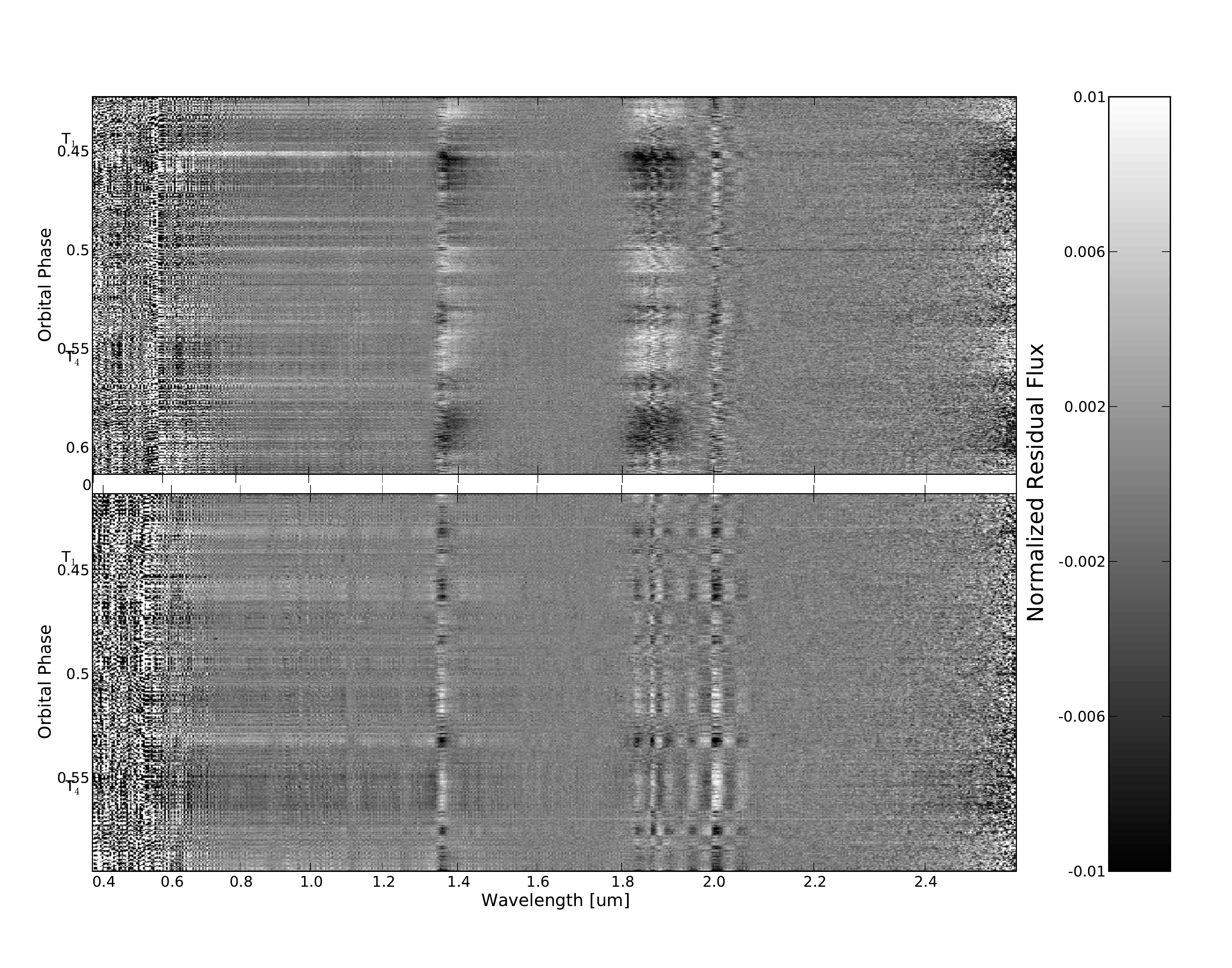}{width=7in}{}{Residuals to the data for the
  nights of 28~Dec ({\em top}) and 30~Dec ({\em bottom}) after fitting
  Eq.~\ref{eq:fluxeqn} to the data in each wavelength channel,
  normalized by the median flux value in each wavelength
  channel. Residual correlated errors remain at shorter wavelengths
  and in regions of strong telluric absorption. The first ($T_1$) and
  fourth ($T_4$) points of contact of the eclipse are noted, as
  calculated from the ephemeris of \cite{hebb:2009}.}

\figtwocol{spectrum}{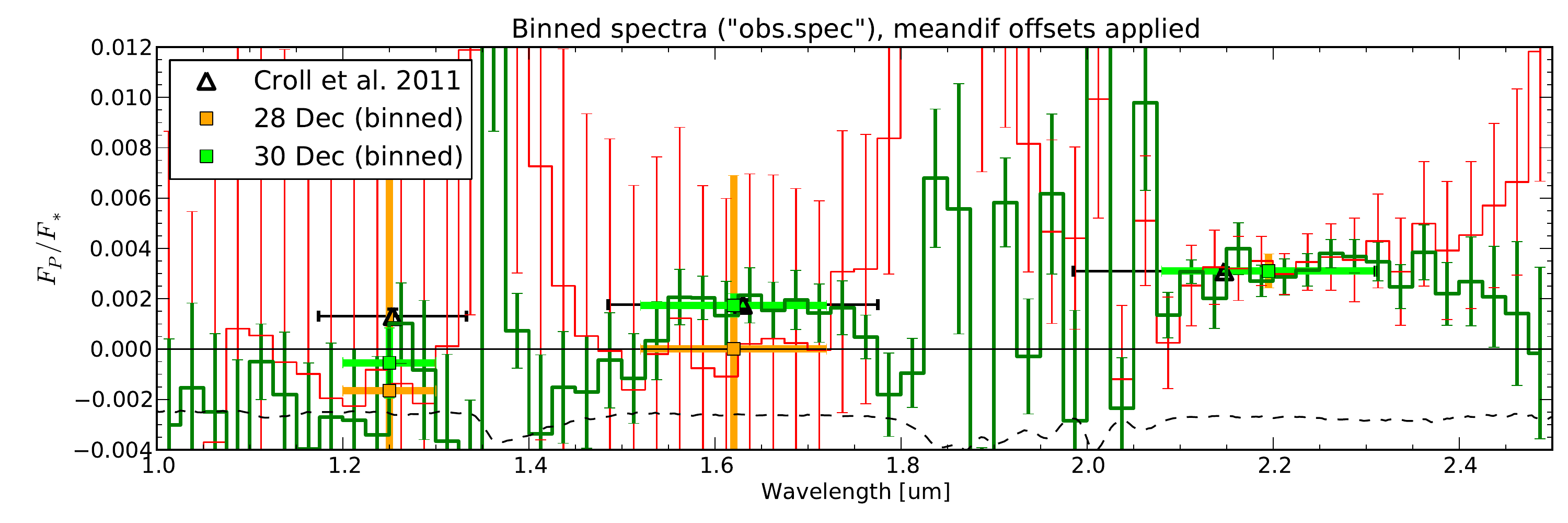}{width=6.5in}{}{Emission spectra (solid
  lines, binned to squares) after calibrating our relative measurement
  with the absolute photometry of \cite{croll:2011a} (black
  triangles).  The 28~Dec observations (red) suffer from extremely
  large systematic uncertainties at shorter wavelengths
  (cf. Fig.~\ref{fig:simulation}), and uncertainties are also large in
  the J band for the 30~Dec observations (green).  The K$-$H color we
  measure on 30~Dec agrees with the previous photometric value.  The
  dashed line at bottom shows the observed telluric extinction
  coefficient. }

Our measurement errors generally increase toward shorter wavelengths
owing to the systematic biases discussed in
Sec.~\ref{sec:atmodisp}. Our performance also worsens in regions of
high telluric absorption; this is either because our simple modeling
does not accurately capture the behavior of saturating absorption
lines, or because the abundances of the absorbing telluric species are
changing with time.  As we describe in Sec.~\ref{sec:telluric} above
we believe the latter description applies to the behavior of the
28~Dec eclipse spectrum in water absorption bands, while the former
applies to the strong \coo\ absorption bands (around $2-2.07$\,\micron
) on both nights.

As expected from Secs.~\ref{sec:atmodisp} and~\ref{sec:allsys} and
Fig.~\ref{fig:simulation}, the large uncertainties for the 28~Dec data
(deriving from our use of a narrow slit and our decision not to guide
along the parallactic angle) prevent the 28~Dec data from usefully
constraining \wtb's emission.  In our final analysis we thus use only
the wide-slit (30~Dec) data, which our modeling suggests are the most
reliable.

\subsection{Comparison With Observations}
As we have noted throughout, we make only a relative eclipse
measurement because division by the slit loss term removes a mean
eclipse signature from all channels.  Precise photometric eclipse
measurements \citep{croll:2011a} allow us to tie our observations to
an absolute scale.  From our investigation of systematic effects in
Sec.~\ref{sec:telluric} we expect our measurements to be robust in
telluric-free regions of the H and K bands, but we expect systematics
to limit our precision at shorter wavelengths and in any region of
strong telluric absorption.

The H and Ks filters used by \cite{croll:2011a} cover part of the
telluric absorption band from $1.78-1.98$\,\micron , and without an
independent calibration source we strongly mistrust our extracted
spectra in these regions.  We average our spectra over wavelength
ranges corresponding approximately to the CFHT/WIRCam filter
responses, but modified as necessary to avoid strong telluric
features where our analysis is compromised: we use ranges of
$1.52-1.72$\,\micron\ and $2.08-2.31$\,\micron\ to correspond to the H
and K bands, respectively.  The regions we avoid have greater telluric
absorption, so these wavelengths contribute relatively less to the
photometric measurements. Though telluric contamination thus precludes
a truly homogeneous comparison between our results and those of
\cite{croll:2011a}, using a blackbody model we estimate that the
different wavelength ranges results in a difference of only 0.01\,\%,
well beneath the precision we demonstrate below.

We compute K$-$H contrast colors (i.e., differential eclipse depths)
on 28~Dec and 30~Dec of \hkcolorfirst\ and \hkcolor, respectively.
The former value has a much larger uncertainty for the reasons
discussed above in Sec.~\ref{sec:allsys}: the 28~Dec observations used
a narrow (1.6'') slit and so are much more susceptible to systematic
errors. We thus discard the 28~Dec spectrum and adopt the 30~Dec
spectrum as our best estimate of \wtb's emission. We thus have a K$-$H
contrast color (\hkcolor) fully consistent with, though of a lower
precision than, the photometric value of $0.133\% \pm 0.022$\,\%\
\citep{croll:2011a}.  \wtb 's broadband NIR emission closely
approximates that of a 3,000~K blackbody
\citep{croll:2011a,madhusudhan:2011}; our contrast color is consistent
with a blackbody of temperature \hktemp, confirming this result.  The
30~Dec K$-$J contrast color is \jkcolor, which is also consistent with
the photometric value of $0.178\% \pm 0.031\%$ \citep{croll:2011a} but
is more uncertain: this large uncertainty exists because chromatic
slit losses could substantially bias our measurement at these shorter
wavelengths even with a 3.0'' slit.  Since the J-band uncertainties
are dominated by the seeing-dependent component of chromatic slit
losses, it may be difficult to improve on the short-wavelength
performance we demonstrate here.

The weighted mean difference between our~H and~K measurements and
those of \cite{croll:2011a} is \meandif, consistent with the offset of
0.215\,\%\ expected from spectral models \citep{madhusudhan:2011}
given the wavelengths used in our initial correction with the
achromatic slit loss time series.  We adjust our relative spectra by
this offset and thus place our measurements on an absolute scale. The
calibrated spectra from each individual night are plotted in
Fig.~\ref{fig:spectrum} and we show our final planet/star contrast
spectrum, plotted over the wavelengths we consider to be uncorrupted
by telluric effects, in Fig.~\ref{fig:goodspec} and list the contrast
ratios in each wavelength bin in Table~\ref{tab:goodspec}.

\figtwocol{goodspec}{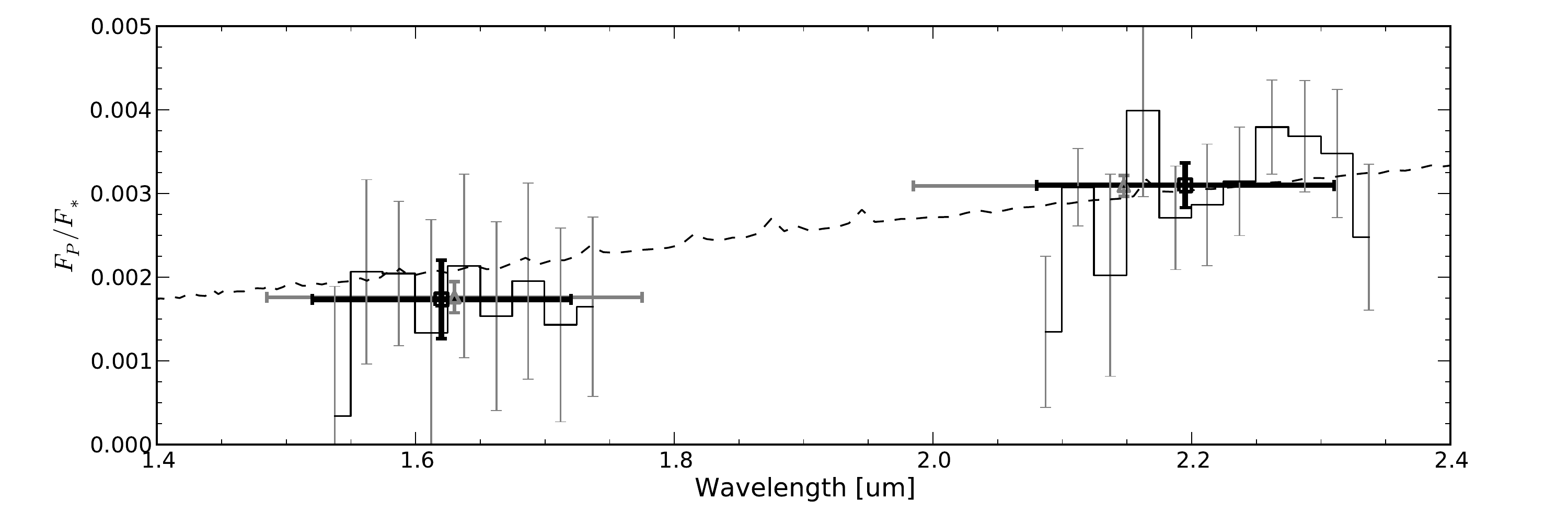}{width=7in}{}{Tentative
  planet/star contrast spectrum of the WASP-12 system from our 30~Dec
  observations.  We have calibrated our relative spectroscopic
  measurements (thin black curve, binned to thick black squares) with
  photometric data points \citep[gray triangles;
  ][]{croll:2011a}. Horizontal errorbars represent the effective
  widths of the photometric bandpasses; the ranges differ to prevent
  telluric variations from corrupting our single-slit
  observations. The dashed line shows the flux expected from a 3,000~K
  blackbody divided by the stellar flux model described in
  Sec.~\ref{sec:energy}.  Systematic uncertainties are larger at
  shorter wavelengths due to chromatic slit losses, so we do not plot
  those data here.}

\begin{deluxetable}{c l }
  \tablecolumns{3} \tablecaption{\label{tab:goodspec} Calibrated
    Planet/Star Contrast Spectrum } 
\tablewidth{0pt}
\tablehead{
\colhead{Wavelength Range (\micron)}  &  \colhead{$F_{P}/F_*$ ($10^{-3}$)\tablenotemark{a}}}
\startdata
$1.525-1.550$ &  0.34  $\pm$  1.55 \\
$1.550-1.575$ &  2.06  $\pm$  1.10 \\
$1.575-1.600$ &  2.04  $\pm$  0.86 \\
$1.600-1.625$ &  1.33  $\pm$  1.35 \\
$1.625-1.650$ &  2.13  $\pm$  1.10 \\
$1.650-1.675$ &  1.53  $\pm$  1.13 \\
$1.675-1.700$ &  1.95  $\pm$  1.17 \\
$1.700-1.725$ &  1.43  $\pm$  1.16 \\
$1.725-1.750$ &  1.65  $\pm$  1.07 \\
$2.075-2.100$ &  1.35  $\pm$  0.90 \\
$2.100-2.125$ &  3.08  $\pm$  0.46 \\
$2.125-2.150$ &  2.02  $\pm$  1.21 \\
$2.150-2.175$ &  3.99  $\pm$  1.03 \\
$2.175-2.200$ &  2.71  $\pm$  0.62 \\
$2.200-2.225$ &  2.86  $\pm$  0.73 \\
$2.225-2.250$ &  3.14  $\pm$  0.65 \\
$2.250-2.275$ &  3.79  $\pm$  0.56 \\
$2.275-2.300$ &  3.68  $\pm$  0.67 \\
$2.300-2.325$ &  3.48  $\pm$  0.77 \\
$2.325-2.350$ &  2.48  $\pm$  0.87 \\
\enddata
\tablenotetext{a}{Quoted uncertainties refer to the relative
  measurements made by our analysis.  The uncertainties of an absolute
  contrast ratio is the quadrature sum of the value listed here and
  \emeandif.}
\end{deluxetable}

\subsection{Spectral Signatures: Still Unconstrained}
The most prominent spectral signature predicted to lie in our spectral
range is the 2.32\,\micron\ \methane\ absorption bandhead
\citep{madhusudhan:2011}.  We consider the model from
\cite{madhusudhan:2011} in which temperature decreases monotonically
with decreasing pressure (the purple curve in their Fig.~1), which is
the model with the largest predicted \methane\ bandhead equivalent
width.  Estimating the continuum using wavelengths from $2.1\,\micron
- 2.25\,\micron$ and measuring the equivalent width from
$2.25\,\micron - 2.34\,\micron$, we calculate this feature's
equivalent width (calculated as a planet/star contrast) to be 16~nm in
the model; with our spectrum we can set a 3$\sigma$ upper limit of
\methanedepth.

We therefore come within a factor of two of being able to measure the
strength of a specific spectral feature, and thus of spectroscopically
constraining atmospheric abundances. However, because the
uncertainties at these wavelengths are dominated by systematics
relating to telluric absorption (cf. Sec.~\ref{sec:telluric}), a
convincing detection would require many eclipses and a more complete
understanding of the effects of telluric methane absorption.

\subsection{Global Planetary Energy Budget}
\label{sec:energy}
In light of our confirmation that \wtb 's NIR contrast color matches
that of a 3,000~K blackbody, we revisit the published eclipse depths
for \wtb\ with an eye toward examining the planet's global energy
budget.  Since the orbital eccentricity is consistent with zero
\citep{campo:2011,husnoo:2011} we neglect tidal effects as a possible
energy source and focus only on reprocessed stellar energy. 

We convert eclipse depths
\citep{lopez-morales:2010,croll:2011a,campo:2011} into surface fluxes
using the known system parameters \citep{hebb:2009}, propagating the
uncertainties in these parameters throughout our subsequent analysis.
We use \cite{castelli:2004} models with $T_{eff} = 6250$ and 6500~K,
$\log g = 4.0$ and ~4.5~(cgs units), and [M/H]$=+0.2$ and~$+0.5$, and
interpolate linearly in each of these quantities to the \wt\
parameters of 6300~K, 4.16, and 0.3.  Using the known \wt\ system
parameters we then convert to planetary fluxes using appropriate
filter transmission profiles in each waveband (z', J, H, and Ks from
the ground, and all four IRAC channels on the Spitzer Space
Telescope)\footnote{We take the effective z' profile to be the product
  of the SPICAM CCD quantum efficiency and the z' filter transmission,
  obtained from the Apache Point Observatory website:
  \url{http://www.apo.nmsu.edu/}.  CFHT/WIRCam profiles are taken from
  the WIRCam website:
  \url{http://www.cfht.hawaii.edu/Instruments/Filters/wircam.html}.
  Spitzer/IRAC filters are the full array spectral response curves,
  found at the IRAC website:
  \url{http://irsa.ipac.caltech.edu/data/SPITZER/docs/irac/calibrationfiles/spectralresponse/}}. In
the same manner we convert our contrast ratio spectrum in
Fig.~\ref{fig:goodspec} into a surface flux spectrum.  We plot the
full set of flux-calibrated eclipse measurements for this system in
Fig.~\ref{fig:fluxspec}.

\figtwocol{fluxspec}{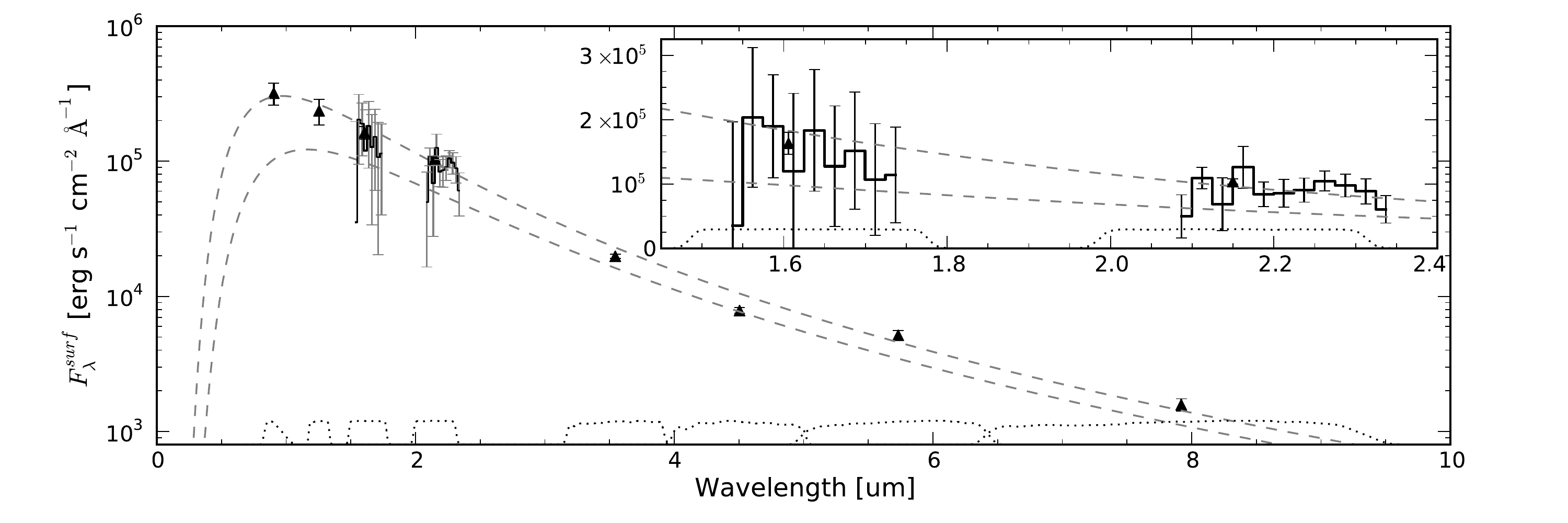}{width=7in}{}{Flux-calibrated
  spectral energy distribution of WASP-12b.  Triangles are from
  previous optical \citep{lopez-morales:2010}, NIR \citep{croll:2011a},
  and mid-infrared \citep{campo:2011} photometry, and the solid line
  (magnified in the inset) shows our measurements. The dashed lines
  show the surface flux of blackbodies with temperatures of 3,000~K
  and 2,500~K. The dotted lines at bottom show the filter transmission
  profiles described in Sec.~\ref{sec:energy}. }

We use the filter profiles and \wtb 's known size to then compute the
dayside luminosity density.  The current set of measurements puts a
lower limit on the dayside luminosity of $2.0\times
10^{30}$~erg~s$^{-1}$. This value is a lower limit because it assumes
zero flux outside the filter bandpasses, which is improbable.
Modeling the planet's spectrum between filters as a piecewise linear
function increases the measured luminosity to $3.6\times
10^{30}$~erg~s$^{-1}$.  As noted previously, the planet's broadband
spectrum closely approximates a 3,000~K blackbody
\citep{madhusudhan:2011}; such a spectrum would emit an additional
$0.7\times 10^{30}$~erg~s$^{-1}$ shortward of the z' band.  The
presence of any optical absorbers \citep[as has been observed on some
planets: e.g.,][]{charbonneau:2002,sing:2011k} would tend to decrease
this optical emission. We thus estimate the planet's total dayside
luminosity to lie in the range $(2.0-4.3) \times
10^{30}$~erg~s$^{-1}$.

On the other hand, \wtb\ absorbs $(1-A_B)(5.1 \pm 0.8) \times
10^{30}$~erg~s$^{-1}$ of stellar energy, where $A_B$ is the planet's
Bond albedo. \cite{cowan:2011} have suggested that the hottest of the
Hot Jupiters (including the 3,000~K \wtb) have low albedos and low
energy recirculation efficiencies; assuming zero albedo, our
calculations limit the nightside luminosity to $(0.8-3.1) \times
10^{30}$~erg~s$^{-1}$.

Approximating the nightside as a blackbody of uniform temperature,
these values correspond to a nightside effective temperature of
2,000$-$2,800~K or a day-night effective temperature contrast of
$200-$1,000~K.  Temperature contrasts of this magnitude would
correspond to a planetary energy recirculation efficiency
\citep{cowan:2011circ} of $\epsilon=2-10$, which suggests that the
planet's recirculation efficiency may not be as low as predicted if
the planet also has a low albedo \citep[cf.][]{cowan:2011}.  This
result should be relatively easy to test, since these temperature
contrasts imply IRAC1~\&~2 phase curve contrasts \citep[$\Delta
F/\langle F \rangle$, cf.][]{cowan:2007} of as much as $0.25$\,\% and
a Ks-band phase curve contrast of $\lesssim$~0.15\,\%.  Warm Spitzer
can easily reach this precision, and the results will help constrain
\wtb 's recirculation efficiency and Bond albedo.  After referral of
this manuscript we became aware of just such a set of IRAC
observations \citep{cowan:2012}. Though of limited precision, these
Spitzer observations support our predictions above and suggest that
\wtb\ has a nonzero albedo and low recirculation efficiency.

Eclipse observations still do not bracket the flux peak of \wtb 's
emission: $F_\lambda$ increases monotonically from 8\,\micron\ (IRAC4)
to 0.9\,\micron\ (z' band).  We therefore strongly encourage efforts
to detect the planet's emission and/or reflection at shorter
wavelengths (e.g., in the I and R bands) to further refine the
planet's albedo and flesh out its energy budget.  The optical
planet/star contrast ratios are challenging ($<0.08$\,\%) but should
be attainable on modest-sized (3-4~m) ground-based telescopes or with
the Hubble Space Telescope (HST).  HST observations at wavelengths
inaccessible from the ground would also help fill the gaps in the
planet's spectral energy distribution and so decrease the uncertainty
in the planet's dayside luminosity.

\section{Lessons  for Future Observations}
\label{sec:future}
Our primary, but ultimately tentative result is \wtb 's K$-$H contrast
color of \hkcolor.  This result is a factor of 2.5 less precise than
that determined by wide-field, relative eclipse photometry
\citep{croll:2011a} with a comparable amount of observing time.
Nonetheless, we are heartened by our ability to self-calibrate out
correlated noise in pursuit of precise relative measurement and to
come within a factor of two of constraining the strengths of specific
molecular features.  This suggests that many repeated observations
with SpeX or similar wide-slit spectrograph might descry spectral
signatures.  However, more progress must be demonstrated for
single-slit observations to be competitive with photometry.  The field
is only now developing the beginnings of an understanding of telluric
effects on such observations \citep{mandell:2011}, and there is still
no consensus explanation for the full set of observations of
\cite{swain:2010}.

Our analysis in Sec.~\ref{sec:sys} demonstrates that telluric
variations can imprint spurious features on our planetary spectrum,
and chromatic slit losses can induce broad spectral
gradients. Certainly, future observations should guide the slit along
the parallactic angle and use as large a slit as possible.  Future
SpeX observations should make use of MORIS \citep{gulbus:2011}, a
camera allowing simultaneous NIR spectroscopy and optical imaging, to
distinguish between telluric transparency effects and instrumental
throughput variations coupled to variable seeing, pointing, and PSF
morphology.  It could be possible to improve future performance by
modeling the evolution of telluric features using high-resolution
spectra, but we suspect this will not be feasible for very
low-resolution observations such as those presented here.  In any
event, we would prefer to eschew telluric modeling and the many
additional complications such analyses must entail. At this point we
cannot say whether the increase in throughput afforded by prism mode
was worth the cost in lower resolution, but we have SpeX echelle data
in hand (and more pending) that may allow us to answer this question.
In any case higher resolution will not substantially improve the
resolution of our final planetary spectrum, because substantial
binning is still required to achieve a useful S/N.

As we stated in Paper~I, we believe that near-infrared, multi-object
spectrographs (MOS) will be the key technology that will enable
detailed spectroscopic studies of exoplanet atmospheres. As the high
precision achieved with optical and NIR MOS units
\citep{bean:2010,bean:2012} demonstrates, these instruments may well
prove transformative for such studies.  Slits can be made large enough
to avoid all pointing error-induced slit loss effects, and
simultaneous spectra of multiple calibrator stars allow all the
advantages enjoyed by relative photometric techniques to be transferred
to the field of spectroscopy. The spectroscopic calibrator stars
provided by a MOS largely eliminates spectral contamination due to
changes in telluric water transmission (i.e., $\Delta\textrm{Tran}$)
since the calibrator observations remove telluric transmission effects
in the same manner as is done in relative transit photometry. This
should largely obviate the need to model evolving airmass extinction
effects.  Because the large majority of multi-object spectrographs
work at wavelengths where $\Delta\textrm{Rad}$ effects are negligible,
with multi-object observations neither telluric radiance nor
transmittance should prove a confounding factor.

However, a large fraction of the currently known transiting systems
will remain off-limits to the multi-object technique. This can result
either from systems which lack comparison stars of adequate brightness
within several arc minutes of the exoplanet host star, or from host
stars that are too bright to observe with the large-aperture
telescopes currently hosting MOS units.  Transiting planets in these
systems can be observed spectroscopically at $\lambda < 1.7\,\micron$
with Hubble/WFC3 \citep[cf.][]{berta:2012}, but spectroscopy in the K
and L bands (where \methane\ and CO bands are prominent) will remain
the domain of ground-based, single-slit spectroscopy for the near
future.

Whatever the observing technique used, we emphasize the importance of
observing multiple transit or eclipse events with ground-based
observations.  There are many subtle confounding factors in such
analyses, and repeated observations are essential to discriminate
between intermittent systematic effects and a true planetary signal.
Many nights of observations would be required with SpeX to build up a
useful spectroscopic S/N for most systems, but it does seem
feasible. Nonetheless even with large-aperture telescopes single-epoch
observations -- including the results we present here -- may well be
treated with some skepticism.

\section{Conclusions}
\label{sec:conclusion} We have presented evidence for a tentative
spectroscopic detection of near-infrared emission from the extremely
Hot Jupiter \wtb.  Our data are compromised by correlated noise:
spectrophotometric variations induced by telluric variations (owing to
changing airmass and telluric abundances) and instrumental
instabilities (caused mainly by fluctuations in the instrument PSF,
but also by atmospheric dispersion) that are largely, but not wholly,
common-mode across our wavelength range.  By removing a common time
series from all our data we self-calibrate and remove much of this
variability, but biases remain. Though this calibration subtracts an
unknown constant offset from our measured spectrum we renormalize
using contemporaneous eclipse photometry \citep{croll:2011a}.

Although we present a possible emission spectrum of the planet in
Fig.~\ref{fig:goodspec}, uncertainties are still too large (by a
factor of two) to constrain the existence of putative \methane\
absorption features.  We measure a K$-$H contrast color of \hkcolor,
consistent with a blackbody of temperature \hktemp; thus our results
agree with (but are less precise than) previous photometric
observations \citep{croll:2011a}. Our spectroscopic precision is
limited by residual correlated noise and, due to our lack of external
calibrators, by our extreme susceptibility to interference from
telluric and instrumental sources outside a fairly narrow wavelength
range.  Modeling (described in Sec.~\ref{sec:sys}) gives us confidence
that within these regions our planetary spectrum is free of telluric
contamination and (with a 3.0'' slit) chromatic slit losses play a
negligible role at $\lambda > 1.4\,\micron$.

Our primary result is methodological: to avoid biases from chromatic
slit losses single-slit, NIR spectroscopy of transiting exoplanets
should use slits as wide as possible and always keep the slit aligned
to the parallactic angle.  Instruments must be kept well-focused
throughout such observations to minimize the effects of
seeing variations. Substantially more attention must be paid to
telluric variations if observations are to extend beyond the fairly
narrow windows we describe in Sec.~\ref{sec:telluric}.

We predict that multi-object spectrographs will easily achieve better
performance than what we have demonstrated here: wider slits and
multiple simultaneous calibration stars will measure and remove
instrumental and telluric systematics.  These instruments are deployed
on an ever-growing number of large-aperture telescopes and are
beginning to be put to the test.  In the meantime, we hope our
descriptions of these first stumbling efforts will inform future
studies so that the routine, detailed characterization of
exoatmospheres can begin in earnest.

\section*{Acknowledgements}
We thank S.~Bus and J.~Rayner for assistance in preparing and
executing our observations, the IRTF Observatory for supporting our
stay at Hale Pohaku, and the entire SpeX team for that rare
achievement: a superb instrument that's a pleasure to use. Thanks also
to A.~Mandell for discussions about LBLRTM, to D.~Feldman for his
MATLAB wrapper scripts, to M.~Swain for discussions about SpeX, and to
S.~Frewen, B.~Croll, and N.~Cowan for comments during manuscript
preparation.  Finally, we heartily thank our anonymous referee for
insightful comments that demonstrably improved the quality of this
paper.

IC and BH are supported by NASA through awards issued by JPL/Caltech
and the Space Telescope Science Center.  TB is supported by NASA
Origins grant NNX10AH31G to Lowell Observatory.  This research has
made use of the Exoplanet Orbit Database at
\url{http://www.exoplanets.org}, the Extrasolar Planet Encyclopedia
Explorer at \url{http://www.exoplanet.eu}, and free and open-source
software provided by the Python, SciPy, and Matplotlib communities.
Despite the lack of an IRTF public data archive we will gladly
distribute our raw data products to interested parties.

\footnotesize


\clearpage

\clearpage


\end{document}